%% file: main_arxiv.tex
\pdfoutput=1
\documentclass[runningheads]{llncs}

\usepackage[T1]{fontenc}
\usepackage[utf8]{inputenc}
\usepackage{amsmath}
\usepackage{amssymb}
\usepackage{mathtools}
\usepackage{graphicx}
\usepackage[dvipsnames]{xcolor}
\usepackage[normalem]{ulem}

\usepackage{booktabs}
\usepackage{float}
\usepackage{placeins}
\usepackage{tikz}
\usetikzlibrary{arrows.meta,positioning,decorations.pathreplacing}
\usetikzlibrary{patterns}
\usepackage{pgfplots}
\pgfplotsset{compat=1.18}
\usepgfplotslibrary{groupplots}
\pgfplotsset{discard if not/.style 2 args={
  x filter/.append code={
    \edef\tempa{\thisrow{#1}}\edef\tempb{#2}%
    \ifx\tempa\tempb\else\fi
  }}}

\usepackage[ruled,vlined,linesnumbered,noend]{algorithm2e}
\SetAlFnt{\fontsize{8}{8.5}\selectfont}
\SetAlCapFnt{\small}
\SetAlCapNameFnt{\small\bfseries}
\SetAlgorithmName{Pseudocode}{Pseudocode}{List of Pseudocode Blocks}
\SetAlgoNlRelativeSize{-1}

\SetNlSty{ProtocolNl}{}{}
\SetKwBlock{RecordRule}{Record:}{end}
\SetKwBlock{VoteRule}{Vote:}{end}
\SetKwBlock{ResolveRule}{Resolve:}{end}
\SetKwBlock{VoteCertificateRule}{Vote certificate:}{end}
\SetKwBlock{FinalRule}{Final:}{end}
\SetKwBlock{SkipRule}{Skip:}{end}
\SetKwBlock{DecideRule}{Decide:}{end}
\SetKwProg{When}{when}{:}{end}
\SetKw{KwChoose}{choose}
\SetKw{KwBroadcast}{broadcast}
\SetKw{KwRecord}{record}
\SetKw{KwIgnore}{ignore}
\SetKw{KwReconstruct}{reconstruct}
\SetKw{KwForm}{form}
\SetKw{KwSign}{sign}
\SetKw{KwDecide}{decide}
\DontPrintSemicolon
\newcommand{\VC}{\mathsf{VC}}
\newcommand{\SC}{\mathsf{SC}}
\newcommand{\FC}{\mathsf{FC}}
\newcommand{\GST}{\mathrm{GST}}
\newcommand{\pk}{\mathsf{pk}}
\newcommand{\sk}{\mathsf{sk}}
\newcommand{\Msg}[1]{\mathtt{#1}}

\usepackage[appendix=append,bibliography=common]{apxproof}

\newtheoremrep{lemma}{Lemma}

\newtheoremrep{theorem}{Theorem}

\newtheoremrep{corollary}{Corollary}
\spnewtheorem{applemma}{Lemma}[section]{\bfseries}{\itshape}
\spnewtheorem{apptheorem}{Theorem}[section]{\bfseries}{\itshape}
\spnewtheorem{appdefinition}{Definition}[section]{\bfseries}{\itshape}

\usepackage[nocompress]{cite}
\usepackage{hyperref}
\hypersetup{colorlinks=true,allcolors=.}
\apptocmd{\appendixprelim}{%
  \crefalias{section}{appendix}%
  \crefalias{subsection}{subappendix}%
  \crefalias{subsubsection}{subsubappendix}}{}{}
\usepackage[capitalise,nameinlink]{cleveref}
\crefname{algocf}{Pseudocode}{Pseudocode}
\crefname{algorithm}{Pseudocode}{Pseudocode}
\crefname{applemma}{Lemma}{Lemmas}
\crefname{apptheorem}{Theorem}{Theorems}
\crefname{appdefinition}{Definition}{Definitions}

\newcommand{\Prob}{\Pr}

\DeclareMathOperator*{\argmax}{arg\,max}

\newcommand{\protocolname}{%
  \texorpdfstring{\textsc{ML-FVS}}{ML-FVS}}

\begin{document}
\interfootnotelinepenalty=10000
\raggedbottom

\title{Beyond Leaderless: Randomized Consensus Against Mobile Crashes}
\titlerunning{Beyond Leaderless}
\author{Ittai~Abraham\inst{1} \and Kaya~Alpturer\inst{2} \and
  Pranav~Garimidi\inst{1} \and Joachim~Neu\inst{3} \and
  Tim~Roughgarden\inst{1,4} \and Ertem~Nusret~Tas\inst{1}}
\authorrunning{I. Abraham et al.}
\institute{a16z frontier, \email{\{iabraham,pgarimidi,ntas\}@a16z.com}
  \and
  Princeton University, \email{kalpturer@princeton.edu}
  \and
  Technion, \email{joachimneu@gmail.com}
  \and
  Institute for Advanced Study, \email{tim.roughgarden@gmail.com}}
\hypersetup{pdftitle={Beyond Leaderless: Randomized Consensus Against Mobile Crashes},
  pdfauthor={Ittai Abraham, Kaya Alpturer, Pranav Garimidi,
  Joachim Neu, Tim Roughgarden, Ertem Nusret Tas}}
\maketitle

\begin{abstract}
  \input{abstract}
\end{abstract}

\setbox0=\vbox{\tableofcontents}

\input{sections/introduction}
\input{sections/model}
\input{sections/protocol}

\input{sections/adaptive}
\input{sections/rational}
\input{sections/evaluation}
\nosectionappendix
\begin{toappendix}
\input{sections/timing-appendix}
\input{sections/evaluation-appendix}
\end{toappendix}

\subsubsection*{Acknowledgments.}
This work was done while the second author was interning at a16z frontier research.

\subsubsection*{AI disclosure.}
Generative AI tools, such as GPT-6 Astra, were used during the research
process and preparation of this manuscript. The authors take full
responsibility for its contents.

\bibliographystyle{splncs04}
\bibliography{refs}

\nosectionappendix
\begin{toappendix}
\input{sections/related}
\input{sections/comparison-attacks}
\input{sections/cadence-attack}
\end{toappendix}

\end{document}

%% file: abstract.tex
Many consensus protocols fail to terminate under
attack from an adversary that has the power to
move between replicas and repeatedly crash one of them, even in synchrony.
We present Multi-Leader Fixed-View Simplex (\protocolname{})
to defend against such an adversary while keeping the simple structure
and good-case latency of Simplex consensus.
Our protocol extends fixed-view Simplex with multiple leaders,
each of which may broadcast a proposal.
At the voting deadline, each replica votes on the set of proposals
it has seen so far.
A hidden threshold beacon revealed after the votes gives a priority
order on the multiple leaders,
resolving each vote to its highest-priority proposal.
Our model combines permanent adaptive Byzantine faults with mobile crashes.
For a constant crash budget and $\Delta$-delayed Byzantine corruption,
we obtain $O(\Delta)$ expected latency over repeated views whose leaders
start honest.
With immediate Byzantine corruptions, the protocol
still achieves liveness.
We also evaluate the protocol under random network delays and stochastic
faults, finding that multiple leaders reduce view failures even when rational
leaders delay their proposals to increase rewards.

%% file: sections/introduction.tex
\section{Introduction}
\label{sec:introduction}

A well-known lower bound is that in a synchronous network,
consensus tolerating $t$ crashes requires at least $t+1$
communication rounds in the worst case~\cite{lf82crashrounds,at99bivalency}.
An even stronger setting is one in which a \emph{mobile} adversary
is able to observe the execution and change which replica crashes.
Santoro and Widmayer~\cite{sw89healer,as24mobileflp} showed that
even one mobile crash can lead to an infinite execution of a deterministic
consensus protocol under synchrony.\footnote{These classical lower bounds
assume agreement and strong unanimity/validity. The first bound assumes $n>t+1$.}
Randomized protocols, on the other hand, can avoid this issue by
making sure non-terminating executions have probability zero.
The key problem to avoid is that, if liveness depends on a single,
publicly known leader, the adversary can attack that leader
and crash it, moving on to its replacement in the following view.
Simply rotating leaders does not resolve this problem when the
adversary can follow the rotation.

Several protocols in the literature reduce reliance on a single leader,
but their guarantees against mobile faults differ
(see \cref{sec:related} for further comparisons):

In \emph{Leaderless Consensus}, Antoniadis, Desjardins,
Gramoli, Guerraoui, and Zablotchi~\cite{adggz21leaderless}
study resilience to
\emph{mobile suspension}\footnote{Note that in a suspension (sometimes also called \emph{atomic crash}),
the replica does not send any message in the prescribed period
while a (regular, non-atomic) crash can interrupt a broadcast and deliver a subset of messages.}:
in their message-passing model,
the suspended replica stops listening and replying for a single
round of two message delays while the adversary may
select a different replica in the next round.
Their BFT-Archipelago protocol tolerates $f-1$
Byzantine replicas and one suspended replica among
$n=3f+1$ replicas, with an $O(n)$-round worst-case
latency bound, too high for practical consensus.

Cadence~\cite{babel26cadence} shows that the difference between
mobile suspension and mobile crashes matters for protocols for
multiple proposers since, unlike a suspension, a crash can
partially deliver messages.
One way to interpret their protocol is that it implements a virtual
leader via forming a quorum over the success or failure of each
individual proposer.
While a suspended proposer can be excluded consistently in the views of replicas using a quorum
certificate, a crash during the proposal broadcast can split
positive and negative votes such that
neither inclusion nor exclusion receives a quorum.
Then, the view fails and Cadence must start
a new agreement instance for the same slot to recover.

Mysticeti~\cite{bck24mysticeti} orders blocks through a public sequence of
leaders. Leaders earlier in this sequence must be committed or skipped before
later blocks can be output. While a quorum can skip a suspended proposer,
as in the attack on Cadence, a crash during the proposal broadcast can leave only some
replicas with the proposal. The proposal may then receive too few votes to be
committed and too few negative votes to be skipped. Later proposals can then
have all their commit certificates and still wait for a decision on an earlier
leader.

An alternative approach is VABA~\cite{ams19vaba}, which
runs $n$ parallel
leader-based protocols and then selects one using shared randomness.
This prevents the adversary from knowing in advance which
instance will determine the outcome.
VABA provides this guarantee in asynchrony, but its high latency
even in the good case makes it impractical.
Our aim is to use this delayed selection with fewer leaders and
lower latency under partial synchrony.
\begin{samepage}
    We thus ask:
    \begin{quote}
        \textbf{Can we tolerate a small number of adaptive
        mobile crashes while keeping the simple structure
        and fast good case of Simplex?}
    \end{quote}
\end{samepage}

Our protocol, \protocolname{}, extends
fixed-view Simplex~\cite{dt2026simplexfvs,cp23simplex}.
Each view has $k$ publicly designated leaders, each of which may
broadcast a proposal.
The protocol keeps the proposal,
voting, and finalization steps of Simplex and ranks the
leaders
at random after voting.
We achieve this by making replicas vote on the set of proposals
that they have seen and include shares
of a $(2f+1)$-threshold randomness beacon.
When votes are combined, the beacon reveals a random
priority order,
and each vote resolves to its highest-priority
proposal.

We consider adaptively chosen permanent Byzantine
corruptions and temporary crashes.
Both count against a total fault budget $f$, while crashes
also count against a separate crash budget $c$. A
Byzantine corruption reveals the replica's state
and secrets and once activated never refunds its budget.
A crash preserves that state and reveals no secrets.
Healing a crash resumes the replica immediately,
but the crash counts against both budgets for another $\Delta$.
Message delays are bounded by $\Delta$ after $\GST$.
We consider both immediate Byzantine corruption,
which takes effect as soon as its target is chosen (delay $D=0$),
and corruption that takes effect $\Delta$ after its target is
chosen (delay $D=\Delta$).
Mobile crashes are always immediate.

With $\Delta$-delayed Byzantine corruption, we call a
view starting at least $\Delta$ after $\GST$
a \emph{good-case view} if all its leaders
are honest
at the view's start and no earlier Byzantine attacks are pending against them.
In this good case, mobile crashes are still
allowed, and
the adversary may start new Byzantine attacks during the
view.

At the start of a view, the adversary can interrupt proposal
broadcasts without knowing which leader the beacon will
rank first.
In the good case, the view succeeds if that leader
completes
its broadcast even if the others do not.
Note that crashing the highest-priority leader later
cannot erase already sent messages.
For instance, in a good-case view with two leaders and one
mobile crash, at least one leader completes its proposal broadcast.
Since the beacon ranks such a leader first with
probability at least $1/2$,
each such view produces a final certificate within
$3\Delta$ with probability at least $1/2$. Repeated good-case views give
expected latency $6\Delta$.
\begin{theorem}[Main result, informal]
    \label{thm:informal-main}
    With $n=3f+1$ replicas, $k=c+1$
    leaders suffice against $c$ mobile
    crashes:
        each good-case view finalizes a block within $3\Delta$ with probability
        at
    least $1/(c+1)$, even with $f-c$ Byzantine replicas.
    Repeated good-case views
        give expected latency at most $3\Delta(c+1)$.

    \textbf{In particular, two leaders tolerate one
    mobile
    crash, requiring at
            most two views on average.}
\end{theorem}

Moreover, with randomly sampled leaders,
the protocol still  achieves liveness even when some leaders are
Byzantine, under both immediate and delayed Byzantine corruption
(\cref{sec:adaptive}).

\subsection{Main Contributions}
\label{sec:contributions}

\begin{itemize}
    \item \textbf{Safety.}
          We prove
          that honest replicas running \protocolname{} never finalize
          conflicting blocks under mobile crashes and adaptive Byzantine corruption
          using quorum intersection
          (\cref{sec:protocol}).
    \item \textbf{Liveness under adaptive faults.}
          With uniformly randomly sampled $c+1$ leaders,
          final certificates are produced in infinitely many views
          under both immediate and
          delayed Byzantine corruption (\cref{sec:adaptive}).
          In addition to the good-case result above,
          with $\Delta$-delayed
          Byzantine corruption,
          we also prove constant expected
          latency for constant $c$ mobile crashes
          and $c+1$ leaders if leaders are announced
          at the start of the view.
    \item \textbf{View failures under stochastic faults.}
          We evaluate how multiple leaders reduce view failures
          under random network delays and stochastic leader stalls.
          We measure the probability that a view fails to finalize as we vary
          the number of leaders.
    \item \textbf{Rational proposal timing.}
          We analyze leaders that delay proposals to collect more
          valuable transactions~\cite{sst23time}.
          We characterize equilibrium proposal delays and evaluate
          whether multiple leaders still reduce view failures at equilibrium.
          In our evaluation, adding
          a second leader reduces the computed
          view-failure rate from about
          $1\%$ to about $0.01\%$ at equilibrium
          (\cref{sec:rational}).
\end{itemize}

%% file: sections/model.tex
\section{Model}
\label{sec:model}

We study our protocol {\protocolname{}} in a partially
synchronous network against an adaptive
adversary with permanent Byzantine faults and mobile
crashes.

\paragraph{Setup.}
Let $n=3f+1$ be the number of replicas, where $f$ is the total fault budget
defined below. Our protocol will use a quorum size of $Q=n-f=2f+1$.
Every message is signed by its sender.
We assume a public
external-validity function that checks a block and its
proof~\cite{ckps01broadcast}. A block is \emph{externally valid}
if this function accepts.
We assume that an honest leader with the
required certificates can construct an externally valid extension of the block chain, and its proof.

Replicas have synchronized clocks and $\Delta>0$ is known. The parameter
$\Delta$ determines the fixed view schedule in \cref{sec:protocol}. Messages
delivered by a scheduled time are processed before the protocol actions,
and messages generated by those actions are processed after.

\paragraph{Leader selection.}
For $1\le k\le n$, each view $v$ has a uniformly sampled set
of $k$ distinct leaders.
Leader sets are sampled independently across views, using randomness
independent of the threshold beacon that later determines their priority order.
The selected replicas are indexed by
$L_v[j]$, $j\in[k]$.
The schedule is public by each view's start
and may be announced in advance.

\subsection{Adaptive Byzantine Corruptions and Mobile Crashes}
\label{sec:partial-synchrony}
A replica is \emph{honest} until a permanent Byzantine
corruption takes effect.
After corruption, the adversary controls its behavior
and may read, copy, or modify its local state,
including signing keys, beacon shares, and protocol records.
An \emph{always honest} replica is one that never becomes Byzantine.
An honest
replica is \emph{active} unless currently crashed.
A crashed replica neither sends nor processes messages, but preserves
its local state and secrets without revealing them to the adversary.
Messages arriving while it is crashed are buffered and immediately delivered
and processed upon healing.
Initially all replicas are honest and active.

\begin{definition}[Adaptive mobile adversary]
    \label{def:adaptive-mobile}
    Fix a total fault budget $f$, a mobile-crash budget
    $0\le c\le f$, a permanent
        corruption delay $D\in\{0,\Delta\}$, and a budget refund delay $\tau=\Delta$.\footnote{We treat $c$ as constant and independent of $n$.}
    The remaining budgets are
    initially $F=f$ and $C=c$. The adversary
    has the
        following operations.\footnote{A corruption or crash can target only
        an active honest replica with no pending permanent corruption.
        A healed replica may be targeted again before a prior crash refund,
        but the new operation uses more budget. Each healing schedules its
        refunds exactly once, and later operations do not change those refunds.}
    \begin{enumerate}
        \item \textbf{Permanent Byzantine corruption.} If
        $F>0$, the adversary may
              start a corruption of an active honest
              replica at time $t$. The target
                  is fixed then, and $F$ decreases
                  immediately by one and $C$ is
                  unchanged. The target becomes Byzantine at $t+D$.
              The replica remains Byzantine forever, cannot
              be healed, and the budget consumed by this
                  corruption is never refunded.
        \item \textbf{Mobile crash.} If $F>0$ and $C>0$,
        the adversary may crash an
              active honest replica. The crash takes effect
              immediately and decreases
              both $F$ and $C$ by one.
        \item \textbf{Healing.} The adversary may heal a currently crashed
        replica
              at time $t$. The replica immediately resumes
              honest execution with its
              preserved state. Neither budget is restored at
              healing. At $t+\tau$, both
              $F$ and $C$ increase by one.
    \end{enumerate}
\end{definition}

At most $f$ distinct replicas ever
become Byzantine, and at most $c$ crashes
    are awaiting a budget refund at a time. These crashes, together with all
    started permanent corruptions, total at most $f$. The
    adversary observes
the public execution and Byzantine replicas' states. If a crash interrupts a
broadcast,
it can choose which messages were sent before the
interruption.

\paragraph{Timing.}
We assume that messages sent
between distinct replicas take positive time.
After the unknown global stabilization time
$\GST$, messages sent by active honest replicas arrive within $\Delta$.
Before $\GST$, delays have no known bound, but messages in transit at $\GST$ arrive
by $\GST+\Delta$.
An honest recipient receives or buffers the message regardless of
whether it is crashed or active when the message arrives.
A message sent by an active replica
cannot be erased by later faults.
Healing resumes interrupted broadcasts to the remaining recipients, with
delays measured from each send.

\subsection{Threshold Priority Beacon}
\label{sec:beacon-model}
Each view uses a threshold beacon, revealed during voting, to
order
    its public leaders.

\begin{definition}[Threshold beacon]
    \label{def:threshold-beacon}
    Setup supplies
    public parameters $\pk$,
        including share verification keys,
        and one private beacon share $\sk_i$ for each
        replica.
        For a tag $T$, the operations are:
    \begin{itemize}
        \item $\mathsf{Share}(\sk_i,T)$ produces replica
        $i$'s share $\sigma_{i,T}$.
        \item $\mathsf{VerifyShare}(\pk,i,T,\sigma_{i,T})$
        checks a share
                  using the corresponding share verification key.
        \item $\mathsf{Combine}(\pk,T,S)$ accepts $Q=2f+1$
        valid shares from distinct
              replicas for that tag, and produces a permutation
              $\pi_T$ of replica identities with proof $\eta_T$.
        \item $\mathsf{VerifyOutput}(\pk,T,\pi_T,\eta_T)$
        checks the output and proof.
    \end{itemize}
    We require correctness and
    uniqueness: every verified
    output for a fixed tag is the same. Before enough
    shares are available to reconstruct, an output must remain pseudorandom
    given the public execution
    and the
    adaptively revealed states of at most $f$ permanently
    corrupted replicas over the whole
    execution. In the ideal analysis the outputs are fresh independent
    uniform permutations. A cryptographic implementation gives computational
    security with an error bound.
\end{definition}

View $v$ uses the tag
$T_v=(v,\mathtt{priority})$\footnote{We
consider a single protocol execution. If keys are reused across executions,
messages and beacon tags must also include a session identifier.}
and reconstructs
    the priority permutation $\pi_v$ with proof $\eta_v$.
    The permutation's restriction to the $k$ leaders is
    uniform.
    In particular,
the priority order remains uniform given the execution up to
the proposal step
    and the keys revealed by Byzantine corruption
    before the voting deadline.

\paragraph{Release and reconstruction.}
Let $s_v$ denote the start time of view $v$.
At $s_v+\Delta$, every active honest replica broadcasts its beacon share.
A replica uses $\mathsf{Combine}$ on $Q$ verified shares from distinct
replicas to reconstruct the priority permutation $\pi_v$ and its proof $\eta_v$.
Permutations and proofs received from other replicas are checked
using $\mathsf{VerifyOutput}$.

When $Q$ honest replicas complete their
share broadcasts at the voting deadline after $\GST$, a
recipient that remains active and honest throughout
$(s_v+\Delta,s_v+2\Delta]$ can reconstruct by
$s_v+2\Delta$.
The adversary
    may reconstruct the output as soon as shares are released at
    $s_v+\Delta$.
    \Cref{sec:adaptive} proves that enough honest replicas complete these
    broadcasts
    despite mobile crashes.

We give a reference construction in \cref{sec:beacon-construction}.

\nosectionappendix
\begin{toappendix}
\input{sections/beacon-construction}
\end{toappendix}

%% file: sections/beacon-construction.tex
\section[Threshold Beacon Construction]{Threshold Beacon Construction}
\label{sec:beacon-construction}

Consider threshold BLS with hashed
output~\cite{drand-specification}. Let
$\mathsf{e}:G_1\times G_2\to G_T$ be a nondegenerate pairing of prime-order
groups of order $q>n$, with generator $g_2\in G_2$. Trusted setup, or a
separately
analyzed distributed key generation (DKG) protocol, supplies shares
$x_i=p(\alpha_i)$ of a random polynomial over
$\mathbb F_q$ of degree at most $2f$, at distinct nonzero public points $\alpha_i$,
conditioned on $x=p(0)\ne0$. Replica $i$'s private beacon share is $\sk_i=x_i$.
Publish $Y=g_2^x$ and $V_i=g_2^{x_i}$, included in the public parameters $\pk$.
Use
separate hashes $H_1$ to $G_1$ and $H_2$ to bit strings.
Replica $i$ produces
    $\sigma_i=H_1(T)^{x_i}$; its share check is
\[
    \mathsf{e}(\sigma_i,g_2)=\mathsf{e}(H_1(T),V_i).
\]
For $Q$ verified shares indexed by $I$,
interpolate at zero:
\[
    \lambda_i=\prod_{j\in I\setminus\{i\}}\frac{-\alpha_j}{\alpha_i-\alpha_j},
    \qquad \sigma=\prod_{i\in I}\sigma_i^{\lambda_i}=H_1(T)^x.
\]
A fixed public permutation sampler uses $H_2(T,\sigma)$ as
its seed to produce $\pi_T$. Return this permutation with proof $\eta_T=\sigma$.
To verify a received permutation, check that it matches this seed and that
$\mathsf{e}(\sigma,g_2)=\mathsf{e}(H_1(T),Y)$. Validate group membership and
    canonical encodings before these checks. Every set of $Q$ verified shares for the same tag
    yields the
    same
        $\sigma$. Cryptographic unpredictability relies on
        appropriate BLS hardness
        and random-oracle assumptions and the setup's security.

%% file: sections/protocol.tex
\section{Protocol and Safety}
\label{sec:protocol}
\begin{figure}[t]
    \centering
    \begin{tikzpicture}[x=0.029\textwidth, y=1cm,
            event/.style={align=center, font=\footnotesize}]
        \draw[-{Stealth[length=2mm]}, semithick] (-0.6,0) -- (31.2,0);
        \foreach \x/\lab in {0/{$s_v$}, 10/{$s_v+\Delta$},
        20/{$s_v+2\Delta$}, 30/{$s_v+3\Delta$}}
        \draw (\x,0.10) -- (\x,-0.10) node[below=1pt] {\lab};
        \draw[decorate, decoration={brace, mirror, amplitude=4pt}]
        (0,-0.55) -- (10,-0.55)
        node[midway, below=6pt, event] {\textbf{Proposal window}};
        \node[event, anchor=south] at (10,0.25)
        {\textbf{Vote}\\proposals and shares};
        \node[event, anchor=south] at (20,0.25)
        {\textbf{Skip}\\if no vote
        certificate};
        \node[event, anchor=south] at (30,0.25)
        {\textbf{Next view}};
    \end{tikzpicture}
    \caption{One view of \protocolname{}.
        After $\GST$, replicas that remain active and honest
        reconstruct the beacon by $s_v+2\Delta$. The next
        view starts
        at $s_v+3\Delta$.}
    \label{fig:timeline}
\end{figure}
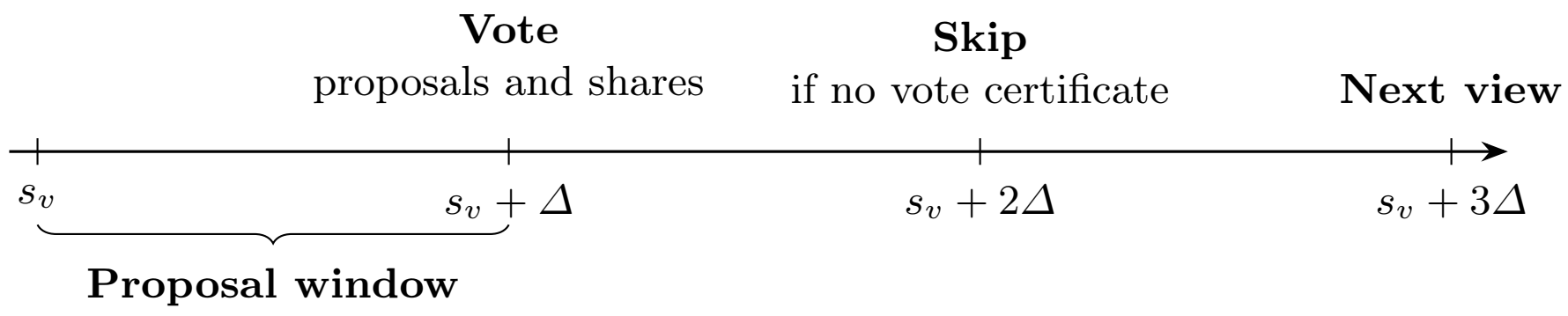

\protocolname{} extends fixed-view Simplex
to multiple leaders. Views are
    numbered from $1$. Each view $v$ starts at $s_v=3v\Delta$
and ends at $s_v+3\Delta$.
Its $k$ distinct leaders are
publicly known by the view start.
Each replica votes on a set of proposals it received by $s_v+\Delta$.
    Replicas verify and broadcast
every certificate when they first form or receive it. At $s_v+3\Delta$, they
enter view $v+1$, sign no further messages
in view $v$, and continue processing and forwarding certificates from earlier
views.
\Cref{fig:timeline} illustrates the view schedule.

\paragraph{Proposal rule.}
\Cref{prot:proposal} gives the rule for each leader.
A valid proposal $\langle\Msg{Propose},v,j,b,w\rangle$ from
leader $L_v[j]$ includes an externally valid
block $b$ of consensus payload, the view number $v$, and
a reference $w$, which is either an earlier view number or a special value denoting genesis.
A vote certificate $\VC_w(b')$ records quorum support for block $b'$
in view $w$, and a skip certificate $\SC_i$ records quorum votes to skip view $i$.
Their formation rules are given below.
For an earlier view $w<v$, the proposal
includes $\VC_w(b')$ for a block $b'$ that $b$ extends, and $\SC_i$ for every
view $w<i<v$.
For genesis, $b$ extends genesis and the proposal includes
$\SC_i$ for every prior view $i<v$. At $s_v$, each
leader $j$ proposes using
the highest
$w$ for which it has the required certificates, treating genesis as lower than every
view. We call such a reference justified.
It sends nothing if no such reference exists.

\begin{algorithm}[tbp]
    \caption{Proposal rule for leader $j$ in view $v$.}
    \label{prot:proposal}
    \When{the local clock reaches $s_v$}{
        \eIf{some $w$ is justified}{
            \KwChoose the highest justified $w$\;
            \KwBroadcast a valid proposal
            $\langle\Msg{Propose},v,j,b,w\rangle$\; }{
            send nothing\;
        }
    }
\end{algorithm}

\begin{figure}[tb]
    \centering
    \begin{tikzpicture}[x=1.35cm, y=0.72cm, scale=0.9, transform shape,
        every node/.style={font=\small},
        vote/.style={draw, rounded corners=1.5pt, fill=black!3,
                minimum height=0.55cm, minimum width=1.05cm, inner sep=3pt},
        result/.style={draw, circle, minimum size=0.52cm, inner sep=0pt,
                font=\small\bfseries},
        map/.style={-{Stealth[length=1.6mm]}, semithick}]
        \node at (2.1,2.55) {$\pi_v:\;Y \succ X \succ Z$};
        \node[font=\scriptsize, anchor=east] at (-0.65,1.35) {vote};
        \node[font=\scriptsize, anchor=east] at (-0.65,0) {resolves to};
        \node[vote] (v1) at (0,1.35) {$\{X,Y\}$};
        \node[vote] (v2) at (1.4,1.35) {$\{X,Y,Z\}$};
        \node[vote] (v3) at (2.8,1.35) {$\{Y,Z\}$};
        \node[vote] (v4) at (4.2,1.35) {$\{X\}$};
        \node[result] (r1) at (0,0) {$Y$};
        \node[result] (r2) at (1.4,0) {$Y$};
        \node[result] (r3) at (2.8,0) {$Y$};
        \node[result] (r4) at (4.2,0) {$X$};
        \draw[map] (v1) -- (r1);
        \draw[map] (v2) -- (r2);
        \draw[map] (v3) -- (r3);
        \draw[map] (v4) -- (r4);
        \draw[semithick] (-0.35,-0.55) -- (-0.35,-0.75) --
        (3.15,-0.75) -- (3.15,-0.55);
        \node at (1.4,-1.15) {$Q=3\;\Longrightarrow\;\VC_v(b_Y)$};
    \end{tikzpicture}
    \caption{Example vote resolution for $Q=3$. Under $Y\succ X\succ Z$, three
        votes resolve to $Y$ and form $\VC_v(b_Y)$.}
    \label{fig:resolution}
\end{figure}
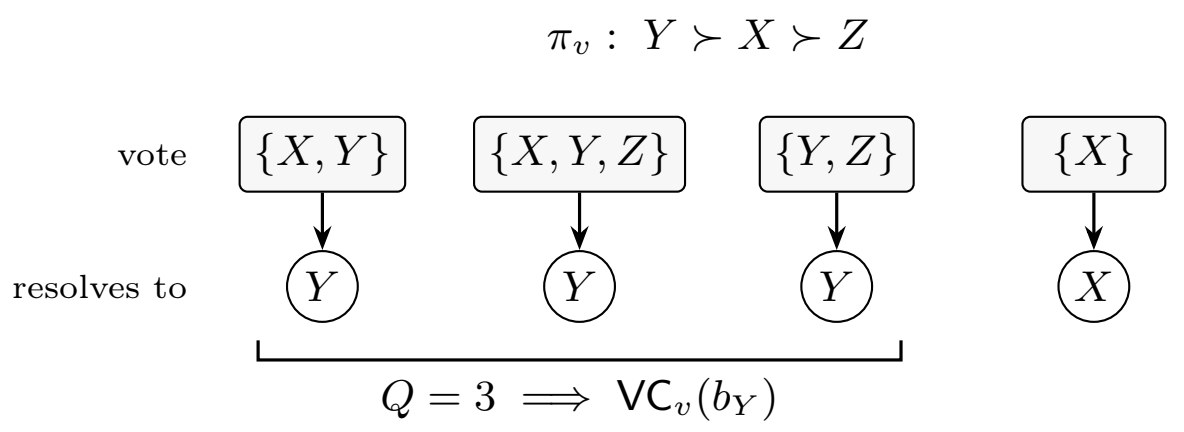

\paragraph{Voting rule and beacon reveal.}
\Cref{prot:voting} gives the voting rule.
At time $s_v+\Delta$, each active honest replica $i$ sends
a vote $\langle\Msg{Vote},v,R_i,\sigma_{i,v}\rangle$.
    Here $R_i$ is its set of recorded
    valid proposals for view $v$, and
$\sigma_{i,v}=\mathsf{Share}(\sk_i,T_v)$ is its beacon
share. After verifying shares for tag
    $T_v=(v,\mathtt{priority})$ and reconstructing the beacon
    order
from $Q$ valid shares as in \cref{sec:beacon-model}, each valid
vote resolves to the (leader, block) pair of its highest-priority proposal,
or to $\bot$ if the set is empty. A vote certificate
consists of $Q$ valid votes from distinct
    replicas that resolve to the same leader and block, together with
    the beacon output and its proof.
    \Cref{fig:resolution} illustrates vote resolution under a priority order.

    Beacon shares are verified with $\mathsf{VerifyShare}$ and counted
    once per replica toward reconstruction.\footnote{A valid share
    can still be used even if the vote contains an invalid proposal or more than
    one proposal from the same leader.}
    A received beacon output is checked against its proof using
    $\mathsf{VerifyOutput}$.
    At the voting deadline, each active honest replica fixes the set
    of proposals included in its vote before processing any shares released at
    that deadline. The set cannot subsequently change.

A valid vote is correctly signed and contains only valid proposals for its view, with at most one per
leader.
An honest replica signs at most one vote per view, with
its single beacon share.

\begin{algorithm}[tbp]
    \caption{Voting rule for proposals in view $v$.}
    \label{prot:voting}
    \RecordRule{
        \When{a valid proposal for view $v$ from
        leader $j$
        is received by
            $s_v+\Delta$, before your vote
            set is fixed}{
            \KwRecord it only if it is the first for
            leader $j$\;
        }
    }
    \VoteRule{
        \When{the local clock reaches $s_v+\Delta$}{
            fix $R$ as the set of
            recorded proposals (possibly $\emptyset$)\;
            \KwBroadcast
            $\langle\Msg{Vote},v,R,\mathsf{Share}(\sk_i,T_v)\rangle$\;
        }
    }
    \ResolveRule{
        \When{$Q$ shares from distinct replicas passing
        $\mathsf{VerifyShare}$ for $T_v$ have been
            collected}{
            $(\pi_v,\eta_v)\gets\mathsf{Combine}(\pk,T_v,S)$,
            for the collected shares $S$\;
            resolve each valid $\Msg{Vote}$, including later arrivals, to the
            highest-priority (leader, block) pair in its
            signed set $R$, or to
            $\bot$ if
            $R=\emptyset$\;
        }
    }
    \VoteCertificateRule{
        \When{$Q$ valid $\Msg{Vote}$ messages from distinct replicas have been
            collected and resolve to the same pair with block $b$}{
            \KwForm $\VC_v(b)$\;
        }
    }
\end{algorithm}

\paragraph{Finalization and skip rules.}
\Cref{prot:final-skip} gives the finalization and skip rules.
An active honest replica that obtains $\VC_v(b)$ signs and
broadcasts a finalization vote $\langle\Msg{Final},v,b\rangle$ if it has signed neither
a finalization vote nor a skip vote in view $v$.
At
$s_v+2\Delta$, it signs and
broadcasts a skip vote $\langle\Msg{Skip},v\rangle$
if it has no vote certificate and has
signed neither a finalization vote nor a skip vote. A final
    certificate $\FC_v(b)$ consists of $Q$ finalization votes for
$b$; a skip certificate $\SC_v$
consists of $Q$ skip votes. Each certificate requires
signatures from distinct replicas for the same view.
Replicas
form certificates whenever the required signatures arrive.
A vote certificate and a skip certificate may both exist.
When an honest replica obtains $\FC_v(b)$, it finalizes $b$ and its
ancestors. It appends the newly finalized blocks to its output log, with each
block following its parent.

\begin{algorithm}[tbp]
    \caption{Voting rule for finalization and skip votes in view $v$.}
    \label{prot:final-skip}
    \FinalRule{
        \When{$\VC_v(b)$ is obtained}{
            \If{you signed neither $\Msg{Final}$ nor $\Msg{Skip}$ in view $v$}{
                \KwSign and \KwBroadcast $\langle\Msg{Final},v,b\rangle$\;
            }
        }
    }
    \SkipRule{
        \When{the local clock reaches $s_v+2\Delta$, after you have processed
            delivered messages}{
            \If{no $\VC_v(\cdot)$ is known and you signed neither $\Msg{Final}$
            nor
                $\Msg{Skip}$ in view $v$}{
                \KwSign and \KwBroadcast $\langle\Msg{Skip},v\rangle$\;
            }
        }
    }
    \DecideRule{
        \When{$\FC_v(b)$ is obtained}{\KwDecide $b$\;}
    }
\end{algorithm}

\paragraph{Communication.}
Apart from beacon setup and
    certificate forwarding, the scheduled protocol steps use at most $k$
proposal broadcasts, $n$ vote broadcasts, and $n$ finalization or skip broadcasts,
for $O(kn+n^2)$ point-to-point messages
in a view where replicas follow
    these steps.

Each vote contains up to $k$ proposals. If a complete
proposal, including
its validity proof and justifications,
takes at most $P$ bytes, votes use
$O(kn^2P)$ bytes in total. A vote certificate contains $Q$ votes, so
broadcasting one
full vote certificate from every replica can add $O(kn^3P)$ bytes.

\subsection{Safety}
\label{sec:safety}
\label{sec:safety-liveness}
We prove safety regardless of message delays.
Any two quorums intersect in at least $f+1$ replicas. Since
at most $f$ replicas ever become Byzantine, the intersection contains an always
honest signer, even if later corruptions reveal keys for
past views.
We defer proofs to \cref{lem:validity-apx}.

Every vote certificate contains an honest vote, which resolves only
to an externally valid block.
\begin{lemmarep}[External validity]
    \label{lem:validity}
    Every block certified by a vote certificate is externally valid.
\end{lemmarep}
\begin{appendixproof}
    A vote certificate $\VC_v(b)$ contains an always honest
    replica's $\Msg{Vote}$ resolving to $b$.
    Such $\Msg{Vote}$ messages contain only valid proposals, so
    $b$ is externally valid. \qed
\end{appendixproof}

Observe that two vote certificates in the
same view always share an honest
signer, whose vote must resolve to the same leader and block pair.
\begin{lemmarep}[Unique vote certificate]
    \label{lem:unique-vc}
    In any view, vote certificates can certify at most one
    $(\text{leader},\text{block})$ pair.
\end{lemmarep}
\begin{appendixproof}
    Suppose vote certificates
    exist for pairs $(j,b)$ and $(j',b')$ in view $v$. Their
    signer sets must intersect in an always honest replica. An
    always honest replica signs at most one $\Msg{Vote}$ in the
    view, and the common beacon order resolves that signed $\Msg{Vote}$ to at
    most
    one pair. Hence $(j,b)=(j',b')$. \qed
\end{appendixproof}

Similarly, a final certificate and a skip
certificate in the same view
would share an honest signer, who cannot sign both.
\begin{lemmarep}[Final and skip
certificates]
    \label{lem:exclusive}
    In any view, a final certificate and a skip certificate cannot both exist.
\end{lemmarep}
\begin{appendixproof}
    Suppose for a contradiction that both $\FC_v(b)$ and $\SC_v$ exist. Each
    certificate has $Q$ distinct signers, so an always honest replica has signed
    both $\langle\Msg{Final},v,b\rangle$ and $\langle\Msg{Skip},v\rangle$.
    This contradicts the rule that an always honest replica
    signs $\Msg{Final}$ or $\Msg{Skip}$ only if it has signed neither in that
    view.
    \qed
\end{appendixproof}

A finalized view cannot be skipped, so later valid proposals must
extend its finalized block.
\begin{lemmarep}[Agreement]
    \label{lem:agreement}
    At most one block is finalized in each view, and every block finalized in a
    later view extends every block finalized in an earlier view.
\end{lemmarep}
\begin{appendixproof}
    Every final certificate contains an always honest $\Msg{Final}$ signer
    and therefore has a vote certificate. By \cref{lem:unique-vc}, all
    final certificates in a view certify the same block.

    Suppose $b$ is finalized in view $v$. Then $\VC_v(b)$ exists, and
    \cref{lem:exclusive} rules out $\SC_v$. We show by induction on $u\ge v$
    that
    every vote certificate in view $u$ certifies $b$ or a descendant of $b$.
    The case $u=v$ follows from
    \cref{lem:unique-vc}.

    For $u>v$, an always honest signer of a vote certificate has checked the
    proposal's justification. A genesis justification, or a justification from a
    view below $v$, would require $\SC_v$, which does not
        exist. The justification must therefore contain $\VC_w(b')$ for some
        $v\le
        w<u$. By induction, $b'$ is $b$ or a descendant, and the certified
        proposal
    extends $b'$. Every final certificate in a later view thus certifies a
    descendant of $b$. \qed
\end{appendixproof}

%% file: sections/adaptive.tex
\section{Liveness}
\label{sec:adaptive}

We first establish quorum availability and certificate catch-up under
the full fault model. We then analyze liveness with fixed Byzantine faults,
followed by mobile crashes and adaptive Byzantine corruption.
The probability bounds below use the
ideal beacon with independent uniform
    outputs in \cref{def:threshold-beacon}.
We defer proofs to \cref{lem:adaptive-active-quorum-apx}.

The delayed refund means that a crash continues to count against the
fault budget after healing. This leaves a quorum that stays active and honest
throughout each interval of length $\Delta$.

\begin{lemmarep}[Active honest quorums]
    \label{lem:adaptive-active-quorum}
    For every time $t$, at least $Q=n-f$ replicas remain
    active and honest
    throughout $(t,t+\Delta]$,
    including the actions at its end.
\end{lemmarep}
\begin{appendixproof}
    If a replica is crashed at some time
    $u\in(t,t+\Delta]$, either it remains
        crashed at the end of the interval or it heals at some $h\ge u>t$.
        If it heals, its charge remains until $h+\Delta>t+\Delta$. Thus a crash
        charge
        for every such replica is outstanding at the end of the interval. Add
        the
        permanent charges of replicas Byzantine by that time. These charges
        total at most $f$, so at most $f$ replicas
        have these charges. Every
    other replica is active and honest throughout the
    stated interval.
        A replica that heals at $t$ may be in this quorum: its charge can be
        refunded
        at $t+\Delta$, but it is active throughout $(t,t+\Delta]$. \qed
\end{appendixproof}

Between voting and skipping, this quorum either forwards a vote
certificate or produces enough skip votes to form a skip certificate.
Since messages are buffered for crashed replicas,
they are also able to obtain these once they are healed.
The following lemma extends the certificate catch-up argument
of \cite{dt2026simplexfvs}.

\begin{lemmarep}[Certificate catch-up]
    \label{lem:certificate-catch-up}
    For every view $w$, any honest replica active at time
    $t\ge\max\{\GST,s_w+2\Delta\}+\Delta$ has either
    $\VC_w(b)$ for some block $b$, or $\SC_w$.
\end{lemmarep}
\begin{appendixproof}
    Fix a view $w$. By \cref{lem:adaptive-active-quorum}, a quorum
    $H$ stays active and honest throughout $(s_w+\Delta,s_w+2\Delta]$.
    Either one member forwards a vote certificate, or all members broadcast
    skip votes.

    Assume $f\ge1$, as the case $f=0$ is immediate.
    Reconstructing the beacon requires another honest replica's share,
    so no honest replica obtains a vote certificate at or before $s_w+\Delta$.
    If a member of $H$ obtains one by $s_w+2\Delta$, it therefore completes
    its broadcast during this interval.
    Otherwise, no member has signed a finalization vote, so all broadcast
    skip votes at $s_w+2\Delta$.

    The certificate or $Q$ skip votes arrive by
    $\max\{\GST,s_w+2\Delta\}+\Delta$ and are processed on arrival or upon
    healing. Hence every honest replica active at or after this time has
    a vote certificate or a skip certificate.
    \qed
\end{appendixproof}

Hence, at each view start $s_v\ge\GST+\Delta$, active
honest leaders have the certificates needed to make justified proposals.

\subsection{Liveness with Fixed Byzantine Faults}
\label{sec:baseline-liveness}

We now consider a set of at most $f$ replicas that are Byzantine
from the start, with no further corruptions or crashes.
We establish liveness when the
highest-priority leader is honest.

\begin{lemmarep}[Liveness with fixed Byzantine faults]
    \label{lem:liveness}
    With a fixed set of at most $f$ Byzantine replicas and
    no crashes, every view $v$ with
    $s_v\ge\GST+\Delta$ whose highest-priority
    leader is honest, produces a final certificate for its block at
    every honest
    replica by $s_v+3\Delta$.
\end{lemmarep}
\begin{appendixproof}
    For every earlier view $u<v$,
    $s_u+2\Delta\le s_v-\Delta$, so
    $\max\{\GST,s_u+2\Delta\}+\Delta\le s_v$.
    By \cref{lem:certificate-catch-up}, every honest
    leader has a vote certificate
    or a skip certificate for every earlier view before its proposal step and
    can therefore
    make a justified proposal.

    Let $j$ be the highest-priority leader. Every honest replica receives
    its valid proposal by the voting deadline and includes it in its
    signed set. The votes and beacon shares of at least
    $Q$
    honest replicas reach every honest replica by $s_v+2\Delta$.
    They reconstruct the same priority output and obtain a
    vote certificate for $j$'s proposal before
    the skip deadline.
        By uniqueness of the vote certificate, any finalization vote already sent in this
        view is
    for the same block. These replicas all broadcast a finalization vote,
    and
        their $Q$ messages arrive by $s_v+3\Delta$,
        forming a final certificate. \qed
\end{appendixproof}

Infinitely many views with
an honest
highest-priority leader give
infinitely many final
certificates. If honest
leaders propose externally valid proper extensions,
agreement then
gives an
    unbounded finalized chain.

\subsection{Liveness with Mobile Crashes}
\label{sec:adaptive-progress}

We now study liveness under mobile crashes,
using hidden priorities to derive good-case success probabilities and expected
latency bounds.

\begin{lemmarep}[A completed highest-priority proposal]
    \label{lem:completed-proposal-progress}
    Suppose a post-$\GST$ view's highest-priority
    leader
    completes a valid
        proposal broadcast at $s_v$ and remains honest at
        every time
        strictly before $s_v+\Delta$. Then an active honest replica obtains
    a final certificate for that proposal by
    $s_v+3\Delta$.
\end{lemmarep}
\begin{appendixproof}
    We follow the proposal through three successive quorums:
    the first votes, the second broadcasts finalization votes, and the third
    receives them.

    If $f=0$, the single replica finalizes at the voting deadline. Assume
    $f\ge1$.
        For $i\in\{0,1,2\}$, let $H_i$ be the replicas active and honest
        throughout
    $(s_v+i\Delta,s_v+(i+1)\Delta]$, including the actions
    at its end.
    By \cref{lem:adaptive-active-quorum}, $|H_i|\ge Q$.

    Every replica in $H_0$ receives or retains the proposal by
    the voting deadline.
    The leader cannot equivocate before the voting deadline. Even if
    it becomes Byzantine
        exactly at the voting deadline, a conflicting proposal sent then cannot
        reach another replica before its vote set is fixed. Each replica in
        $H_0$
    therefore includes the same proposal, and
    its vote resolves to that proposal.

    Later faults cannot erase the sent messages or change the
    signed vote sets.

    Every replica in $H_1$ receives the $Q$
    matching votes and beacon shares from $H_0$
    by $s_v+2\Delta$, reconstructs the
        priority order, and obtains the matching vote certificate
        before
        the skip deadline. An honest replica cannot sign a finalization vote at or before the voting deadline:
        reconstruction
        requires another honest replica's share, first released at the voting deadline and
        received
        after positive transmission time. Nor can it sign a skip vote before the
        scheduled
        skip step. Crashes preserve this signing state. Any earlier finalization vote is
        for the
    same block by \cref{lem:unique-vc}. Thus all replicas
    in
    $H_1$ broadcast a finalization vote at times in
        $(s_v+\Delta,s_v+2\Delta]$.

    Every replica in $H_1\cap H_2$ is active and honest
    throughout $(s_v+\Delta,s_v+3\Delta]$ and receives these
        finalization votes by $s_v+3\Delta$. Since $|H_1\cap H_2|\ge2Q-n=f+1$, an active
        honest
    replica obtains a final certificate by the view's end. \qed
\end{appendixproof}

\begin{lemmarep}[Liveness when leaders remain honest]
    \label{lem:adaptive-honest-leaders}
    Suppose a view $v$ with
    $s_v\ge\GST+\Delta$ has $k>c$
    honest leaders, and no leader
        may become Byzantine during that view, regardless of the
        priority outcome.
    Then an active honest replica obtains a final certificate by $s_v+3\Delta$ with probability at least $(k-c)/k$.
    This holds for both $D=0$ and
    $D=\Delta$.
\end{lemmarep}
\begin{appendixproof}
    After refunds due at $s_v$, any
    crash charge outstanding at or added during
        the proposal step remains until after that step.
        These charges cover at most $c$
    leaders. Every other leader is active, has the
    certificates
    needed to propose by \cref{lem:certificate-catch-up}, and
    completes its valid
        broadcast at $s_v$.
    There are at least $k-c$ such leaders. These
    leaders are fixed before
    the priority beacon is revealed, so
    one ranks first with probability at
    least $(k-c)/k$. On this event,
    \cref{lem:completed-proposal-progress} gives the final
    certificate. \qed
\end{appendixproof}

\paragraph{Good case.}
For $D=\Delta$, a view $v$ with
$s_v\ge\GST+\Delta$ is a
\emph{good-case view} if all its
    leaders are honest at $s_v$ and no permanent attack
    started before
$s_v$ is pending against any of them.
    The adversary may now start new Byzantine attacks during the view, but these cannot take effect before the voting deadline.

\begin{lemmarep}[Good-case liveness]
    \label{lem:delayed-view-progress}
    \label{lem:adaptive-good-case}
    Let $D=\Delta$. In a good-case view with $k>c$
    leaders, an active
    honest replica obtains a final certificate by $s_v+3\Delta$ with probability at least
    $(k-c)/k$.
\end{lemmarep}
\begin{appendixproof}
    By the good-case condition and $D=\Delta$, every
    leader remains honest
    strictly before the voting deadline.
    As in \cref{lem:adaptive-honest-leaders}, at most $c$
    leaders have
        outstanding crash charges or are crashed during the proposal step. At
        least
    $k-c$ leaders therefore complete their broadcasts,
    and
        \cref{lem:completed-proposal-progress} applies when one of these ranks
        first.
    \qed
\end{appendixproof}

With $c+1$ leaders that remain honest until voting, the crash
budget leaves at least one leader able to complete its proposal broadcast.
Because the priority order is hidden until after proposals, the adversary can only guess which leader will rank first.
This gives the two-leader guarantee highlighted in the introduction.
\begin{corollary}[Small leader sets]
    \label{cor:adaptive-two-leaders}
    Under \cref{lem:adaptive-honest-leaders} or
    \cref{lem:adaptive-good-case},
    $k=c+1$ gives success probability at least
    $1/(c+1)$. In particular, two leaders give
    probability at least $1/2$
    with $c=1$, even with $f-1$
    Byzantine replicas outside the set.
\end{corollary}

When the same conditions hold in successive views, each
independently sampled priority order gives another chance to finalize, even after earlier
failures.
The probability of a long run of failures therefore decreases
geometrically, and each additional attempt costs $3\Delta$.

\begin{corollaryrep}[Expected latency]
    \label{cor:protected-expected-progress}
    \label{cor:good-case-expected-progress}
    Let $v_0$ be a post-$\GST$ view chosen using only the history
    available at its start. Suppose the conditions of either
    \cref{lem:adaptive-honest-leaders} or \cref{lem:adaptive-good-case} hold
    in each view from $v_0$ until a final certificate is obtained, with the
    same $k>c$.

        Let $N$ count the views from $v_0$ up to and
        including the first view that produces a final certificate. Then
    \[
        \Prob[N>m]\le(c/k)^m,\qquad
        \mathbb E[N]\le\frac{k}{k-c}.
    \]
    The expected time from $s_{v_0}$ is at
    most
        $3\Delta k/(k-c)$. For $k=c+1$ this is
        $3\Delta(c+1)$, hence $O(\Delta)$
        for constant $c$.
\end{corollaryrep}
\begin{appendixproof}
    Condition on the history at $s_{v_0}$.
    The conditional failure probability in every view is at
    most
        $c/k$. Multiplying these bounds
        gives the tail bound; summing it for $m\ge0$ gives the expectation.
        \qed
\end{appendixproof}

\subsection{Adaptive Byzantine Corruption}

We now consider views whose leaders may already be targeted
for Byzantine corruption. The sampling argument counts leaders outside these
targets that complete their proposal broadcasts at $s_v$ despite crashes.
Since the priority order is uniform and remains hidden during proposals, each
of these leaders has probability $1/k$ of ranking first.
If no new Byzantine corruption starts, these leaders remain honest, so the view
finalizes whenever one of them ranks first. The supporting lemmas and
bounds for general $k>c$ are in \cref{lem:joint-sampling-bound-apx}.

\begin{toappendix}
Let $B_v$ contain the targets of all permanent attacks
started by $s_v$,
    including attacks still pending. Write $b=|B_v|$ and
    $h=|L_v\setminus B_v|$, and let
\[
    d_b=\min\{c,f-b\}.
\]
At most $d_b$ crash charges can be outstanding if no new
permanent attack starts.

\begin{applemma}[Finalization or a new permanent attack]
    \label{lem:adaptive-view-progress}
    Consider a view $v$ with
    $s_v\ge\GST+\Delta$.
    For either corruption delay ($D\in\{0,\Delta\}$), with
    probability at least $\max\{h-d_b,0\}/k$, by the view's end,
    an active honest replica obtains a final certificate
    or the adversary starts a new permanent attack during the view.
\end{applemma}
\begin{proof}
    First consider an adversary that starts no new permanent
    attacks in the
    view.
    At least $\max\{h-d_b,0\}$ leaders outside $B_v$ are free
    of outstanding crash
        charges at the proposal step and complete their
        justified broadcasts by \cref{lem:certificate-catch-up}.
        These leaders remain honest. The priority order
        is uniform
        conditional on their completed broadcasts, so
        \cref{lem:completed-proposal-progress} gives success probability at
        least
        $\max\{h-d_b,0\}/k$.

    For an adversary that may start new attacks, consider a
    second execution
        with the same randomness. It follows the original execution until the
        first
        new permanent attack in the view, omits that
        attack, and then continues with
        any allowed behavior that starts no new permanent attacks. The earlier
        bound applies to the second execution. The executions differ only if the
        original adversary starts a new permanent attack. Otherwise, a final
    certificate obtained in the second execution is also
    obtained in the
        original. Thus success in the second execution implies one of the two
        outcomes in the lemma. \qed
\end{proof}

\Cref{lem:adaptive-view-progress} uses the number $h$ of leaders outside
the targets. When the targets are fixed before the leader set is drawn, we
can average over the draw.

\begin{applemma}[Leader and priority sampling]
    \label{lem:adaptive-sampling}
    Consider a view $v$ with $s_v\ge\GST+\Delta$. Let $B$ contain the
    targets of all permanent attacks started before $L_v$ is drawn, and
    let $b=|B|$.
    For either corruption delay ($D\in\{0,\Delta\}$), with
    probability at least
    \[
        q(b)=\frac{1}{k\binom nk}
        \sum_{h=d_b+1}^{k}(h-d_b)
        \binom{n-b}{h}\binom{b}{k-h},
    \]
    by the view's end, an active honest replica obtains a final certificate
    or the adversary starts a new permanent attack after the draw.
\end{applemma}
\begin{proof}
    First consider an execution with no
    new permanent attacks. The set $B$
    is
        fixed before selection. For $H=|L_v\setminus B|$, we have
    \[
        \Prob[H=h]=\frac{\binom{n-b}{h}\binom{b}{k-h}}{\binom nk}.
    \]
    The adversary may adapt its crashes to the public set,
    but at most $d_b$
        leaders can have outstanding crash charges at the proposal
    step. At least $\max\{H-d_b,0\}$ leaders complete valid
    broadcasts. Their
    identities are fixed before the priority output is
    known, giving probability
        at least $\max\{H-d_b,0\}/k$ that one ranks first. Taking expectation gives
        $q(b)$.
        The comparison of executions in \cref{lem:adaptive-view-progress}, now
        omitting the first new permanent attack after the set draw, proves the
        stated bound when the adversary may start new attacks. \qed
\end{proof}

If the schedule is announced in advance, the adversary may target
leaders after seeing $L_v$; any such attack starts after the draw and so
counts as the second outcome in \cref{lem:adaptive-sampling}. The bound $q(b)$ still depends on the number of targets $b$, which the
adversary chooses. We next show that $q(b)$ is smallest at $b=f-c$, which
gives a bound that holds for every choice of targets.

\end{toappendix}

\begin{lemmarep}[Sampling bound]
    \label{lem:joint-sampling-bound}
    Consider a view $v$ with
    $s_v\ge\GST+\Delta$ and $k=c+1$
    leaders.\footnote{The leader set may be announced in advance.}
    For either corruption delay ($D\in\{0,\Delta\}$), with
    probability at least $(2/3)^k/k$, by the view's end,
    an active honest replica obtains a final certificate
    or the adversary spends a Byzantine corruption.
\end{lemmarep}
\begin{appendixproof}
    We first bound $q(b)$ for every $k>c$. With $B$ and $b$ as in
    \cref{lem:adaptive-sampling} and $H=|L_v\setminus B|$, the bound in
    \cref{lem:adaptive-sampling} is
    \[
        q(b)=\frac{1}{k}\mathbb E\bigl[\max\{H-d_b,0\}\bigr].
    \]
    We show that $q(b)\ge q(f-c)>0$, with
    $q(f-c)\ge(2/3)^k/k$ when $k=c+1$.
    For $b\le f-c$, $d_b=c$, and increasing $b$
    can only
        decrease the number of sampled leaders outside
        $B$. For $b\ge f-c$,
        use the same leader set for two target sets, one obtained by adding a
        target to the other. Adding one target
        reduces $H$ by at most one, while reducing $d_b=f-b$ by exactly one.
        Hence $\max\{H-d_b,0\}$ cannot decrease. This proves the minimum at
        $f-c$. In particular,
    \[
        q(f-c)=\frac{1}{k\binom nk}\sum_{h=c+1}^{k}
        (h-c)\binom{n-f+c}{h}\binom{f-c}{k-h}>0.
    \]
    Positivity follows because $k>c$ and $n-f+c>c$,
    so a uniform leader set has positive probability of containing more than
    $c$ replicas outside a fixed set of $f-c$ targets.
        When $k=c+1$, the positive term requires all $k$
        leaders to avoid $B$.
        Thus
    \[
        q(f-c)=\frac{\binom{n-f+c}{k}}{k\binom nk}.
    \]
    For $0\le j\le c$ and $n=3f+1$,
        $(n-f+c-j)/(n-j)\ge2/3$, which gives the last inequality.
        Combining this bound with \cref{lem:adaptive-sampling} proves the lemma. \qed
\end{appendixproof}

When leaders are drawn at the view's start, delayed
corruptions started at or after the draw cannot take effect before the voting
deadline. The same counting argument then gives probability at least
$(2/3)^k/k$ of finalization in each view, even if new Byzantine corruptions
are started. Repeated attempts give the following expected latency.

\begin{theoremrep}[Expected latency with delayed Byzantine
corruption]
    \label{thm:delayed-fresh-pools}
    Let $D=\Delta$ and $k = c + 1$. Suppose each
    view $v$ with $s_v\ge\GST+\Delta$
    draws its public $k$-element leader set exactly at $s_v$.
    Then the expected time from the start
    of any such view until an active honest replica obtains a final certificate is at most
    $3\Delta k(3/2)^k$, which is $O(\Delta)$ for constant $c$.
\end{theoremrep}
\begin{appendixproof}
    Corruptions started at or after the leader draw cannot take
    effect before voting. We use this to bound the probability of finalization
    in each view, then the expected number of views.

    Fix $k>c$. Let $B$ contain the replicas targeted for Byzantine corruption
    before this view's draw, including pending corruptions, and write $b=|B|$.
     The set $B$ is fixed before selection. Hence the
        distribution of $H=|L_v\setminus B|$ is the one given in
        \cref{lem:adaptive-sampling}. At most $d_b=\min\{c,f-b\}$ leaders can
        be excluded by outstanding crash charges or by crashes during the
        initial
    proposal step. At least $\max\{H-d_b,0\}$ leaders outside
    $B$ therefore complete
        justified broadcasts at $s_v$.

    These leaders remain honest strictly before the voting deadline.
    The adversary does
        not know the priority order when their broadcasts complete. With
    probability at least $\max\{H-d_b,0\}/k$, one of them ranks
    first and
        \cref{lem:completed-proposal-progress} gives a final certificate. Taking expectation
        over $H$ gives $q(b)$ for a final certificate.

    The proof of \cref{lem:joint-sampling-bound} gives
    $q(b)\ge q(f-c)>0$. Set $q_\star=q(f-c)$, and let $N$ count this view and
    each subsequent view through the next final certificate. The success
    probability is at least $q_\star$ conditional on every earlier history.
    Multiplying the conditional failure bounds and summing the tail gives
    \[
        \Prob[N>m]\le(1-q_\star)^m,\qquad
        \mathbb E[N]\le\frac{1}{q_\star}.
    \]
    Thus the expected time is at most $3\Delta/q_\star$.
    For $k=c+1$, the proof of \cref{lem:joint-sampling-bound} gives
    $q_\star\ge(2/3)^k/k$, yielding the stated bound. \qed
\end{appendixproof}

The leader draw in \cref{thm:delayed-fresh-pools} requires
randomness separate from the priority beacon, which is released only at the
voting deadline. Announcing leaders before the view starts would give the
adversary time to begin corrupting them before $s_v$.

The adversary can start at most $f$ Byzantine corruptions,
so eventually no new ones occur. For every fixed set of $f$ corruption targets,
independent uniform sampling produces infinitely many leader sets with more
than $c$ replicas outside that set, with probability one. Hidden priorities then give
repeated chances to finalize, even when corruptions take effect immediately
and the leader schedule is public in advance.

\begin{theoremrep}[Liveness]
    \label{thm:adaptive-hidden-priorities}
    Let $k>c$.
    Active honest replicas obtain final certificates in infinitely
    many views both for $D=0$ and $D=\Delta$.
\end{theoremrep}
\begin{appendixproof}
    Infinitely many views give a probability of at least $1/k$
    of either finalization or a new Byzantine corruption. Only finitely many
    corruptions can occur, so infinitely many of these views must finalize.

    For any fixed set $B$ of at most $f$ replicas,
    an independently sampled
    $k$-element set has positive probability of
    containing more than $c$
    replicas outside $B$, since $n-f>c$ and $k>c$.
    Independence across views
        makes this occur infinitely often with probability one. There are only
        finitely many such sets $B$, so with probability
        one this property holds
        for every one of them.

    Let $B^*$ contain all Byzantine corruption targets, so
    $|B^*|\le f$. The property above applies to $B^*$ even when it depends
    on the public schedule.
    For each view $v$, let $h_v=|L_v\setminus B_v|$, using
    the actual target set
        at its start. Since
        $B_v\subseteq B^*$, every set with more than $c$
    replicas outside $B^*$ has $h_v>c$. Thus infinitely
    many views $v$ with $s_v\ge\GST+\Delta$
    satisfy $h_v>c$. This condition is known before the
    view's priority
        reveal and does not use the future set $B^*$.

    At each such view, \cref{lem:adaptive-view-progress}
    gives conditional
    probability at least $1/k$ that a final certificate is obtained or
    a new permanent
        attack is started.
        The probability of neither outcome in the next $m$ such views is at
        most $(1-1/k)^m$.
        The same tail argument after every fixed view shows that one of
    these two outcomes occurs infinitely often
    with probability one. New
        permanent attacks can be started in only finitely many views, so
    infinitely many must produce a final certificate. \qed
\end{appendixproof}

\subsection{Single Leaders and Known Priorities}
\label{sec:adaptive-assumptions}
We illustrate why the protocol uses multiple leaders and
keeps their priority order hidden until voting.

\begin{lemmarep}[A single leader]
    \label{lem:single-leader-obstruction}
    For $f\ge1$, a mobile adversary with $c=1$ can prevent
    the $k=1$ protocol
    from producing any final certificate, without starting any
    Byzantine corruption.
\end{lemmarep}
\begin{appendixproof}
    At each view start, crash the public
    leader before it
    sends its proposal.
    Heal it at $s_v+\varepsilon$ for
    $0<\varepsilon<\Delta$.
    Even if it immediately resumes its broadcast, delay
    the messages to other replicas
    by $\Delta$, past the voting deadline. At most
    its own vote supports the
    proposal,
    fewer than $Q$, so neither a vote certificate
    nor a final certificate can form.
    The crash charge is refunded against both budgets
    at
        $s_v+\Delta+\varepsilon$, before the next view.
    All replicas are active at the skip deadline and can receive a skip certificate by the next view
    start. This attack can repeat indefinitely. \qed
\end{appendixproof}

Even with multiple leaders, an adversary that knows their priorities
can split the votes and prevent finalization in a view.

\begin{example}[Known priorities]
    \label{ex:adaptive-known-priorities}
    Let $n=4$, $f=c=1$, and $k=2$, with no Byzantine
    replicas. If the priority
    order is known before proposals, crash the highest-priority
    leader $i$
    during
    its broadcast, after the proposal has been sent to itself
    and one other replica.
    Heal it before the voting deadline and delay the
    messages to the remaining recipients by $\Delta$.
    Let the proposal from the other leader $j$
    reach every replica after $i$
    heals
    and before the voting deadline. Two vote sets then
    contain both proposals, and two contain
    only $j$'s proposal. Leader $i$
    gets two resolved votes, as does
    $j$, so neither reaches the quorum of three.
    If the order was instead hidden and uniform
    while this delivery pattern was
        fixed, $j$ would rank first with probability
        $1/2$; all four votes would
        then
    resolve to $j$ and the view would finalize.
\end{example}

%% file: sections/rational.tex
{
\section{Reliability and Proposal Timing}
\label{sec:rational}

We next study whether multiple leaders improve reliability when
leaders can increase their rewards by delaying proposals.
For our evaluation, we use a probabilistic
network model with empirical link-delay
distributions.\footnote{For the empirical distributions, we used
round-trip time (RTT) quantiles from \href{https://www.cloudping.co/}{CloudPing},
accessed on 11 August 2026.}
While this network model captures a realistic simulation setting
as the latencies are derived from real-world measurements,
we also add stochastic faults that model random software and hardware
failures that
can delay proposals~\cite{gmac25ambulance}.

Moreover, observe that
in an implementation of
{\protocolname{}} with $k$ leaders,
the timing of the leaders' actions directly
affects the
outcome of each view. We therefore model each leader as
rational, as timing games
have been studied
theoretically~\cite{sst23time,ktbr26incentives} and observed in
practice~\cite{w24timing}.
We argue that a realistic
common-case evaluation
requires taking them into account.
We do not model adaptive attacks or recovery from mobile crashes.
We use two placements: a global placement
uniform across cities
around the world, and a regional placement
concentrated in North America and
Europe
(\cref{fig:topology}).

We consider one view with $k$ public leaders and an available
priority beacon. Proposals are valid and justified at view start, and replicas
follow the voting rules. Each directed link has a known delay distribution,
from which message delays are sampled independently. Here, $\Delta$ is the
voting deadline measured from view start and need not bound message delays.
Formally, each leader independently stalls with
probability $p$. A stall adds a delay of $\Delta^{\mathrm{stall}}$ to its
proposal to all recipients.
We define the resulting timing game below, then compare
view-failure rates at equilibria.

\input{figures/topology}
\paragraph{Value of delay.}
In many blockchains, delaying a proposal can increase its value because more
transaction fees and maximal extractable value
    (MEV) can be collected.
Following \cite{sst23time,okv23mev}, we
use an increasing reward function
$V(s)=B+M(s)$, where $B$ is the base reward and $M(s)$ is the value accumulated
by proposal time $s$.
Leaders trade off this increase in value against the risk that
their proposals arrive too late to be finalized.

\begin{definition}[Timing game]
    \label{def:timing-game}
    \begin{enumerate}
        \item Each leader $i$ chooses a proposal time
        $s_i\in[0,\Delta]$, where
        $\Delta$ is
              the voting deadline.
        \item Leaders know one another's locations, but
        choose before network
        delays, stalls,
              and the priority order are revealed.
        \item Leader $i$ maximizes its expected
        reward,
        receiving
              $V(s_i)$ if its block is finalized and zero otherwise.
    \end{enumerate}
\end{definition}

The formal equilibrium definition and an analysis with equal
link delays appear in \cref{sec:timing-analysis}.
}

%% file: figures/topology.tex
\begin{figure}[tbp]
  \centering
  \begin{tikzpicture}[
      x=0.028cm,y=0.028cm,
      map land/.style={fill=black!5,draw=black!18,line width=0.15pt,
        line join=round},
      region/.style={circle,fill=black!78,draw=white,line width=0.35pt,
        text=white,font=\tiny\sffamily\bfseries,inner sep=0pt,
        minimum size=8.5pt},
      legend region/.style={region,minimum size=7.5pt},
      legend entry/.style={anchor=west,font=\scriptsize\sffamily,
        text=black!78,inner sep=0pt},
    ]
    \path[use as bounding box] (0,-49) rectangle (350,113);
    \begin{scope}
      \clip (0,0) rectangle (350,113);
      \input{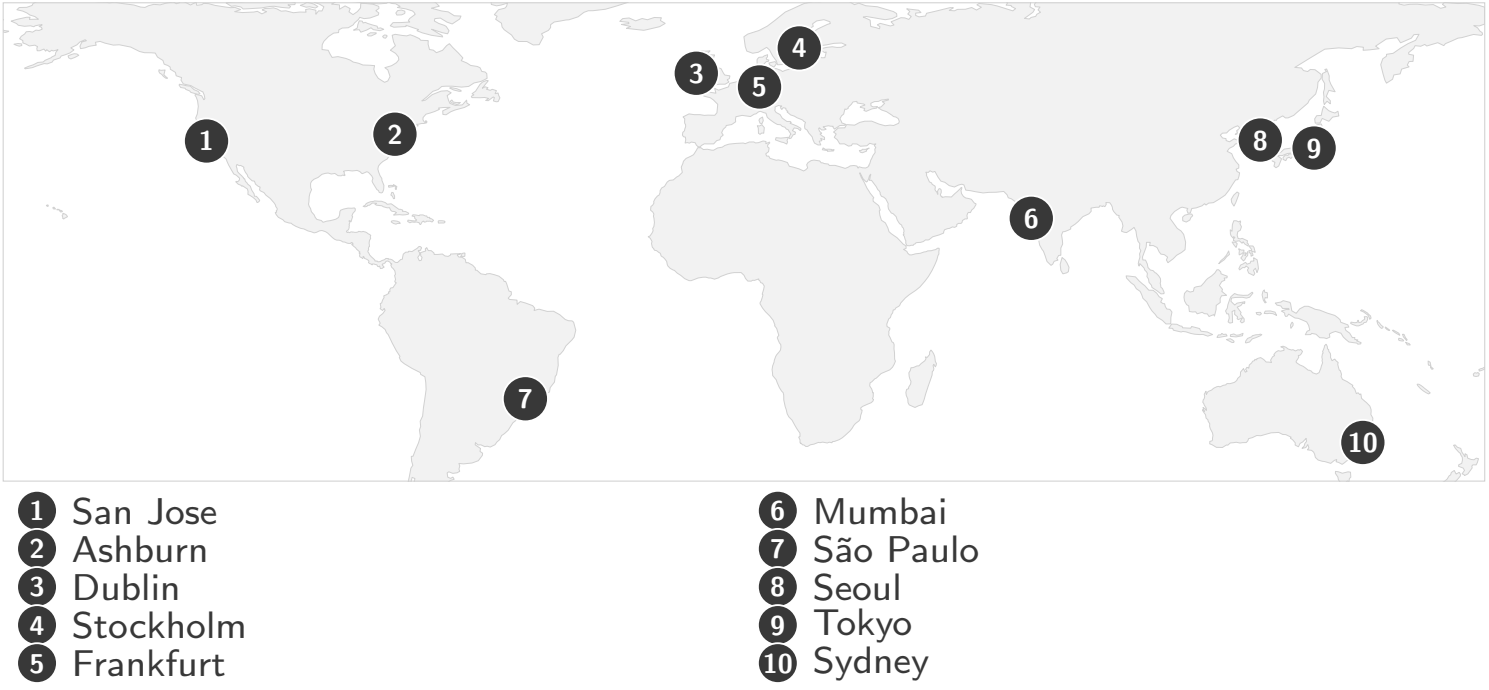}
    \end{scope}
    \draw[black!18,line width=0.2pt] (0,0) rectangle (350,113);
    \node[region] at (48.04,80.35) {1};
    \node[region] at (92.55,81.95) {2};
    \node[region] at (163.74,96.35) {3};
    \node[region] at (188.07,102.33) {4};
    \node[region] at (178.68,93.11) {5};
    \node[region] at (242.88,62.08) {6};
    \node[region] at (123.37,19.45) {7};
    \node[region] at (296.98,80.57) {8};
    \node[region] at (309.69,78.68) {9};
    \node[region] at (321.21,9.13) {10};

    \node[legend region] at (8,-7) {1};
    \node[legend entry] at (16,-7) {San Jose};
    \node[legend region] at (8,-16) {2};
    \node[legend entry] at (16,-16) {Ashburn};
    \node[legend region] at (8,-25) {3};
    \node[legend entry] at (16,-25) {Dublin};
    \node[legend region] at (8,-34) {4};
    \node[legend entry] at (16,-34) {Stockholm};
    \node[legend region] at (8,-43) {5};
    \node[legend entry] at (16,-43) {Frankfurt};

    \node[legend region] at (183,-7) {6};
    \node[legend entry] at (191,-7) {Mumbai};
    \node[legend region] at (183,-16) {7};
    \node[legend entry] at (191,-16) {S\~ao Paulo};
    \node[legend region] at (183,-25) {8};
    \node[legend entry] at (191,-25) {Seoul};
    \node[legend region] at (183,-34) {9};
    \node[legend entry] at (191,-34) {Tokyo};
    \node[legend region] at (183,-43) {10};
    \node[legend entry] at (191,-43) {Sydney};
  \end{tikzpicture}
  \caption{Locations of 100 replicas. The global placement assigns ten to each
  city. The regional placement assigns 40 to North America, 40 to Europe, and 20
  elsewhere.}
  \label{fig:topology}
\end{figure}

%% file: sections/evaluation.tex
\subsection{Evaluation Method}
\label{sec:empirics}

\paragraph{Setup.}
We use the two placements and delay model
above, with
$n=100$, $Q=67$, stall probability $p=1\%$, stall delay
$\Delta^{\mathrm{stall}}=500\,\mathrm{ms}$, and $V(s)=0.041+0.0065s$,
with $s$ in
seconds and rewards in ETH.
The initial value and growth rate follow the
Ethereum parameters in~\cite{ktbr26incentives}.
Unless stated otherwise, $\Delta=4$ seconds.
The $k$ leaders are sampled uniformly without
replacement.\footnote{Our simulation code can be accessed here: \url{https://github.com/kalpturer/beyond-leaderless}.}

\paragraph{View failures.}
The analysis assumes that, whenever the
proposal votes support a vote certificate,
at least $Q=67$ replicas obtain it by
$2\Delta$, and their finalization votes
    reach all replicas by $3\Delta$.
    Under this assumption, a vote certificate leads to a final certificate
and prevents a skip certificate. We therefore measure view failure as the
event that
no vote certificate forms from the proposals received by the voting deadline.
\paragraph{Equilibria.}
Starting from prompt proposals, we update each leader's
proposal time in turn
to a best response on a $0.25\,\mathrm{ms}$ grid, holding the others' times
fixed. We check that no leader can
improve its expected reward by choosing another
grid point. We compute the approximate
    equilibria on this grid.
We vary $\Delta$ from $200\,\mathrm{ms}$ to $4\,\mathrm{s}$ for one and two
leaders, then vary $k$ from one to four at $\Delta=4$
seconds. In both
cases, we enumerate all
leader-location configurations,
weighting each by its probability under uniform sampling.

\subsection{Single and Multiple Leaders}
\label{sec:eval-single-multiple}

At $\Delta=4$ seconds, \cref{fig:miss-k} compares the computed
view-failure rates for one to four leaders in both placements.
Adding a second leader reduces the failure rate from about $1\%$ to about
$0.01\%$, even though leaders continue to delay their proposals at equilibrium.

\input{figures/number-leaders}

More precisely, adding a second leader
reduces
the computed view-failure
rate from $1.0029\%$ to $0.01296\%$
in the global placement.
In the regional placement, it falls from
$1.0012\%$ to $0.01125\%$.

For $k=3,4$,
computed failure rates are $0.003063\%$ and $0.002964\%$
in the global placement, and
$0.001350\%$ and $0.001251\%$
in the regional placement.
To understand the proposal times, consider a single leader.
Its expected
reward initially increases with block value,
decreases when stalled proposals no
longer arrive in time, and drops near the
voting deadline as unstalled proposals also fail to obtain a quorum,
as seen in \cref{fig:single-leader-utility} in \cref{sec:additional-evaluation}.
The computed
equilibria use the single-leader proposal time at
each leader's location for all four leader counts.
On average, leaders leave $107.15\,\mathrm{ms}$
in the global placement
and
$79.55\,\mathrm{ms}$ in the regional placement
between proposing and the voting deadline, well below the
stall duration. The model reports no failures
    for a prompt single leader at this voting deadline
    (\cref{fig:miss-deadline} in \cref{sec:additional-evaluation}).
The remaining failures come from later
proposal times.
In this setting, multiple leaders therefore improve reliability
even when rational leaders leave too little time to absorb a stall.

%% file: figures/number-leaders.tex
\usepgfplotslibrary{groupplots}
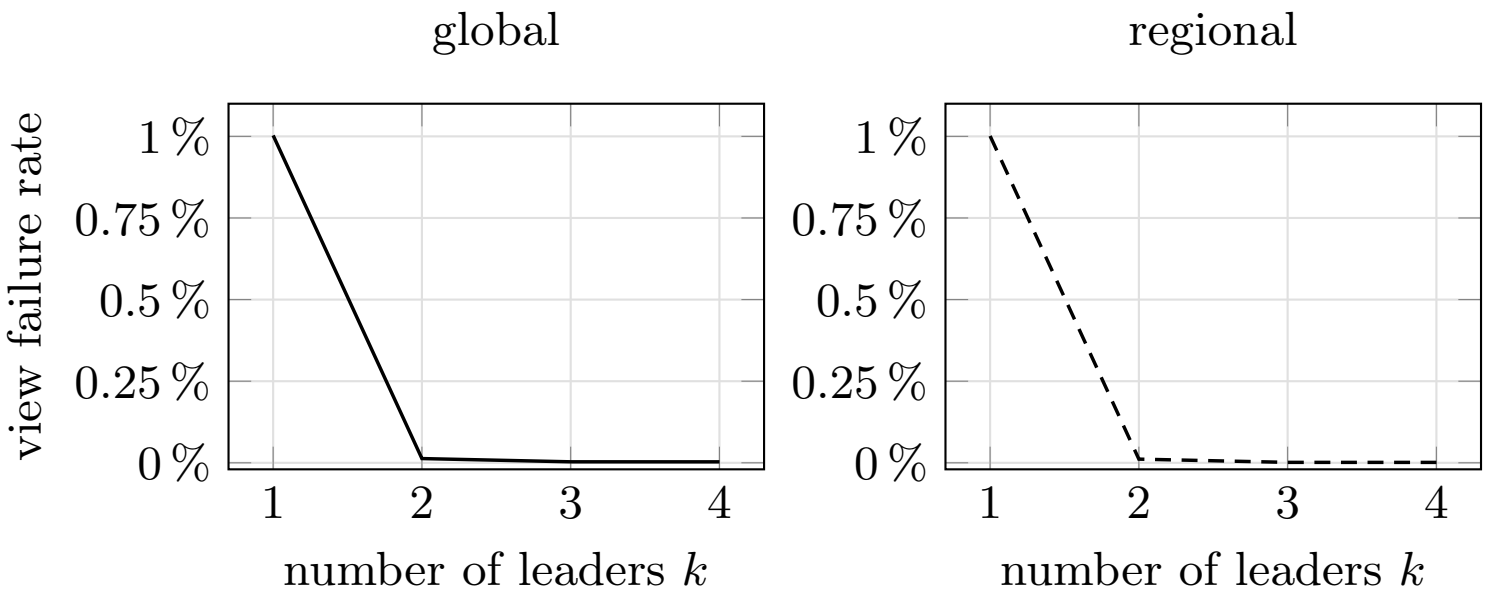
\begin{figure}[tbp]
  \centering
  {
  \begin{tikzpicture}
    \begin{groupplot}[
        group style={group size=2 by 1, horizontal sep=1.2cm},
        width=0.42\linewidth, height=4cm,
        xlabel={number of leaders $k$},
        xmin=0.7, xmax=4.3, xtick={1,2,3,4},
        ymin=-0.0002, ymax=0.011,
        ytick={0,0.0025,0.005,0.0075,0.01},
        yticklabels={$0\,\%$,$0.25\,\%$,$0.5\,\%$,$0.75\,\%$,$1\,\%$},
        scaled y ticks=false,
        ylabel={view failure rate},
        tick label style={font=\footnotesize},
        label style={font=\footnotesize},
        title style={font=\footnotesize},
        grid=major, grid style={black!12},
      ]
      \nextgroupplot[title={global}]
      \addplot[black,line width=0.65pt]
        table[col sep=comma,x=k,y=computed_miss]{data/global-k-stall.csv};
      \nextgroupplot[title={regional}, ylabel={}]
      \addplot[black,densely dashed,line width=0.65pt]
        table[col sep=comma,x=k,y=computed_miss]{data/regional-k-stall.csv};
    \end{groupplot}
  \end{tikzpicture}
  }
  \caption{View failure rates for one to four
  leaders at $\Delta=4$ seconds.}
  \label{fig:miss-k}
\end{figure}

%% file: sections/timing-appendix.tex
\section{Timing Equilibria and Equal Link Delays}
\label{sec:timing-analysis}

\subsection{Leader Timing}
\label{sec:leader-timing}
\label{sec:equilibrium-times}

A single leader's optimal proposal time satisfies {
\[
    s^*\in\argmax_{0\le s\le \Delta}
    V(s)\Prob[\text{the leader's block is finalized}\mid
    s].
\]
}

{
With multiple leaders, let
$U_i(s_i,s_{-i})$ be
leader $i$'s expected reward when
the other leaders use
proposal times
$s_{-i}$.

\begin{appdefinition}[Timing equilibrium]
    \label{def:timing-equilibrium}
    Proposal times $s^*=(s_1^*,\ldots,s_k^*)$
    form a \emph{pure timing equilibrium} if no
    leader can increase its expected
    reward by changing its
    proposal time alone:
    \[
        s_i^*\in\argmax_{a\in[0,\Delta]}U_i(a,s_{-i}^*)
        \qquad\text{for every leader }i.
    \]
\end{appdefinition}
}

\subsection[Equal Link Delays]{Equal Link Delays}
\label{sec:equal-link-delays}

Suppose every link between distinct replicas has the same
fixed delay $\ell>0$, while self-delivery is immediate.
Each leader stalls independently
with probability
$p\in(0,1)$. A stall delays its proposal at every recipient, including its
sender, by $\Delta^{\mathrm{stall}}>0$.

Assume $f\ge1$, $1\le k\le f+1$, and
$\Delta\geq\ell+\Delta^{\mathrm{stall}}$.
The latest proposal times that allow delivery to every replica by the voting deadline, with
and without a stall, are respectively
\[
    s_{\mathrm{safe}}=\Delta-\ell-\Delta^{\mathrm{stall}},
    \qquad s_{\mathrm{late}}=\Delta-\ell.
\]
For sufficiently small $p$, the increase in $V$ gives
$(1-p)V(s_{\mathrm{late}})>V(s_{\mathrm{safe}})$. Otherwise, proposing at
$s_{\mathrm{safe}}$ is optimal, and every view finalizes when all
leaders
choose this time.
\begin{apptheorem}[Equilibrium timing and reliability]
    \label{thm:equal-delay-timing}
    Suppose $V$ is positive and strictly increasing on $[0,\Delta]$, and
    $(1-p)V(s_{\mathrm{late}})>V(s_{\mathrm{safe}})$. With equal link delays,
    the
    unique timing equilibrium has every
    leader propose at
    $s_{\mathrm{late}}$, and the
    probability of no final certificate by view end is $p^k$.
\end{apptheorem}
\begin{proof}
    Call a proposal timely if it reaches every replica by the voting deadline.
    We first show that a view finalizes exactly when at least one
    proposal is timely, then compare the expected rewards from
    $s_{\mathrm{safe}}$ and $s_{\mathrm{late}}$.

    The
    highest-priority timely proposal receives at least $n-(k-1)\ge Q$ votes,
    since only senders of higher-priority proposals that are
    not timely can resolve their
    votes differently. A proposal that
    is not timely receives at most its sender's vote
    and cannot form a certificate. Every replica obtains the resulting vote certificate by
    $\Delta+\ell<2\Delta$ and the final certificate by $\Delta+2\ell<3\Delta$.

    Fix the other leaders' proposal times. Conditional
    on being timely, a
    leader's
    probability of winning is positive and independent of
    its
    own proposal time: stalls are independent, and the beacon selects uniformly
    among timely proposals. Since $V$ is strictly increasing, the only
    possibilities
    for an optimal proposal time are $s_{\mathrm{safe}}$, which is always
    timely,
    and $s_{\mathrm{late}}$, which is timely with probability $1-p$. Later
    proposals cannot win. The two choices give expected
    rewards proportional to $V(s_{\mathrm{safe}})$ and
    $(1-p)V(s_{\mathrm{late}})$, respectively. The inequality makes
    $s_{\mathrm{late}}$ the unique best response, regardless of the other
    leaders'
    choices. At this equilibrium, a view fails exactly when all $k$
    leaders
    stall.
    Independence gives failure probability $p^k$. \qed
\end{proof}

Thus, a second leader reduces the failure probability
from $p$ to $p^2$ without
changing proposal times.

%% file: sections/evaluation-appendix.tex
\clearpage
\section{Additional Evaluation Figures}
\label{sec:additional-evaluation}

\Cref{fig:single-leader-utility} shows the single-leader reward curves.
\Cref{fig:miss-deadline} compares view-failure rates across voting deadlines.

\input{figures/single-leader-utility}
\input{figures/miss-floor}
\FloatBarrier

%% file: figures/single-leader-utility.tex
\usepgfplotslibrary{groupplots}
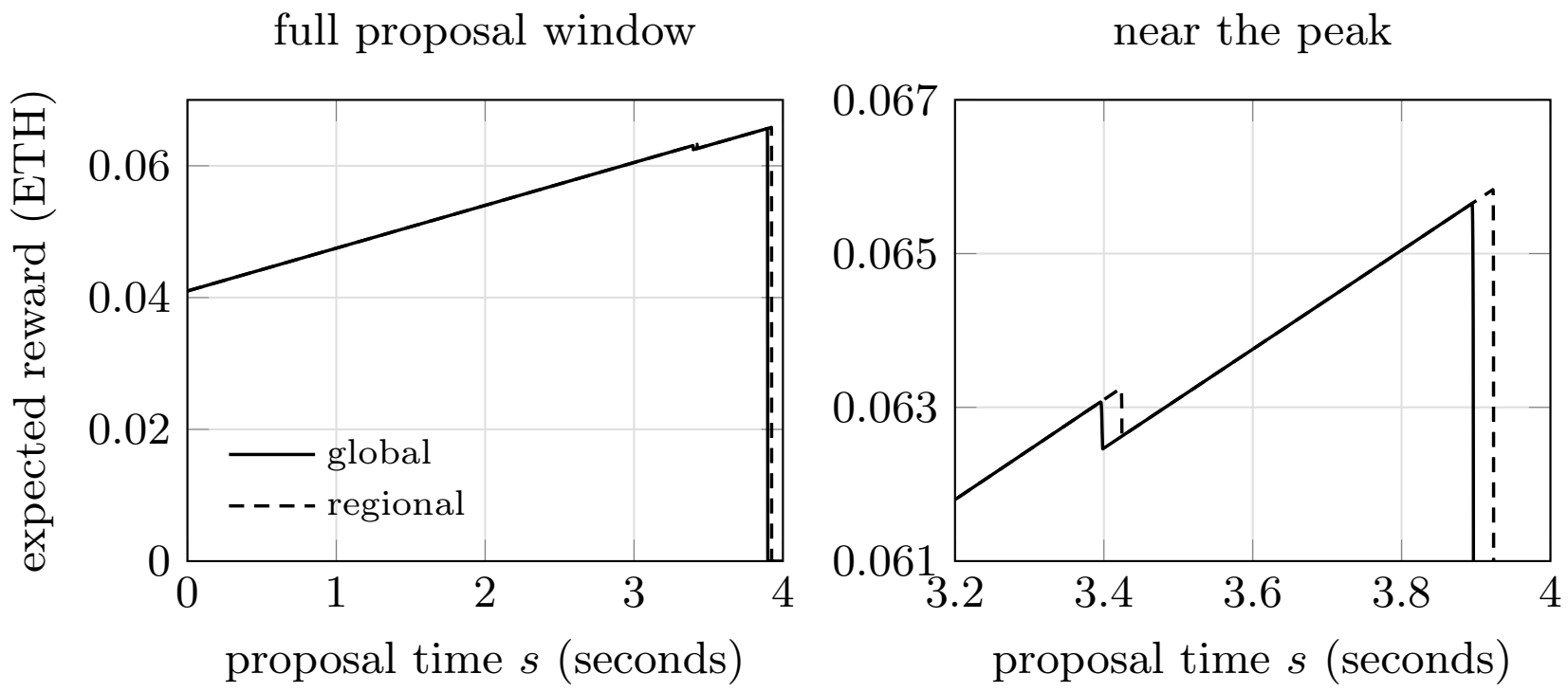
\begin{figure}[!htbp]
  \centering
  \begin{tikzpicture}
    \begin{groupplot}[
        group style={group size=2 by 1, horizontal sep=1.2cm},
        width=0.47\linewidth, height=4.8cm,
        xlabel={proposal time $s$ (seconds)},
        ylabel={expected reward (ETH)},
        scaled y ticks=false,
        tick label style={font=\footnotesize},
        label style={font=\footnotesize},
        title style={font=\footnotesize},
        legend style={font=\scriptsize, draw=none, fill=none,
          at={(0.04,0.04)}, anchor=south west},
        legend cell align=left,
        grid=major, grid style={black!12},
      ]
      \nextgroupplot[
        title={full proposal window},
        xmin=0, xmax=4, ymin=0, ymax=0.07,
        xtick={0,1,2,3,4}, ytick={0,0.02,0.04,0.06},
        yticklabels={$0$,$0.02$,$0.04$,$0.06$}]
      \addplot[black, line width=0.65pt]
        table[col sep=comma,
          x expr={\thisrow{proposal_time_ms}/1000}, y=global_utility_eth]
        {data/single-leader-utility.csv};
      \addlegendentry{global}
      \addplot[black, densely dashed, line width=0.65pt]
        table[col sep=comma,
          x expr={\thisrow{proposal_time_ms}/1000}, y=regional_utility_eth]
        {data/single-leader-utility.csv};
      \addlegendentry{regional}
      \nextgroupplot[
        title={near the peak}, ylabel={},
        xmin=3.2, xmax=4, ymin=0.061, ymax=0.067,
        xtick={3.2,3.4,3.6,3.8,4},
        ytick={0.061,0.063,0.065,0.067},
        yticklabel style={/pgf/number format/.cd,fixed,precision=3}]
      \addplot[black, line width=0.65pt]
        table[col sep=comma,
          x expr={\thisrow{proposal_time_ms}/1000}, y=global_utility_eth]
        {data/single-leader-utility.csv};
      \addplot[black, densely dashed, line width=0.65pt]
        table[col sep=comma,
          x expr={\thisrow{proposal_time_ms}/1000}, y=regional_utility_eth]
        {data/single-leader-utility.csv};
    \end{groupplot}
  \end{tikzpicture}
  \caption{Expected reward for a single
  leader in Frankfurt, with voting deadline
  $\Delta=4$ seconds. The right plot shows the curves near their peaks.}
  \label{fig:single-leader-utility}
\end{figure}

%% file: figures/miss-floor.tex
\usepgfplotslibrary{groupplots}
\begin{figure}[!htbp]
  \centering
  \begin{tikzpicture}
    \begin{groupplot}[
        group style={group size=2 by 1, horizontal sep=1.4cm},
        width=0.47\linewidth, height=4.8cm,
        title style={font=\footnotesize},
        xmode=log,
        xlabel={voting deadline $\Delta$ (seconds)},
        ylabel={view failure rate},
        xmin=0.2, xmax=4,
        ymin=-0.0005, ymax=0.013,
        xtick={0.2,0.5,1,2,4},
        xticklabels={$0.2$,$0.5$,$1$,$2$,$4$},
        ytick={0,0.005,0.01},
        yticklabels={$0\,\%$,$0.5\,\%$,$1\,\%$},
        scaled y ticks=false,
        tick label style={font=\footnotesize},
        label style={font=\footnotesize},
        legend style={font=\scriptsize, draw=none, fill=none},
        legend columns=3,
        legend cell align=left,
        grid=major, grid style={black!12},
      ]
      \nextgroupplot[title={global}, legend to name=missdeadlinelegend]
      \addplot[black!40, dotted, forget plot] coordinates {(0.2,0.01) (4,0.01)};
      \addplot[black, line width=0.65pt]
        table[col sep=comma,
          x expr={\thisrow{deadline_ms}/1000}, y=computed_miss]
        {data/global-k1-stall.csv};
      \addlegendentry{single leader}
      \addplot[black, densely dashed, line width=0.65pt]
        table[col sep=comma,
          x expr={\thisrow{deadline_ms}/1000}, y=computed_miss]
        {data/global-k2-stall.csv};
      \addlegendentry{two leaders}
      \addplot[black, line width=0.75pt,
        dash pattern=on 5pt off 3pt]
        table[col sep=comma,
          x expr={\thisrow{deadline_ms}/1000}, y=prompt_miss]
        {data/global-k1-stall.csv};
      \addlegendentry{single prompt leader}
      \nextgroupplot[title={regional}, ylabel={}]
      \addplot[black!40, dotted, forget plot] coordinates {(0.2,0.01) (4,0.01)};
      \addplot[black, line width=0.65pt]
        table[col sep=comma,
          x expr={\thisrow{deadline_ms}/1000}, y=computed_miss]
        {data/regional-k1-stall.csv};
      \addplot[black, densely dashed, line width=0.65pt]
        table[col sep=comma,
          x expr={\thisrow{deadline_ms}/1000}, y=computed_miss]
        {data/regional-k2-stall.csv};
      \addplot[black, line width=0.75pt,
        dash pattern=on 5pt off 3pt]
        table[col sep=comma,
          x expr={\thisrow{deadline_ms}/1000}, y=prompt_miss]
        {data/regional-k1-stall.csv};
    \end{groupplot}
  \end{tikzpicture}

  \pgfplotslegendfromname{missdeadlinelegend}
  \caption{View failure rates in the global (left) and regional (right)
  placements using the computed proposal times for one and
  two leaders, compared with
  a prompt single leader. Proposals independently stall
  for $500\,\mathrm{ms}$
  with probability $1\%$ (dotted line).}
  \label{fig:miss-deadline}
\end{figure}
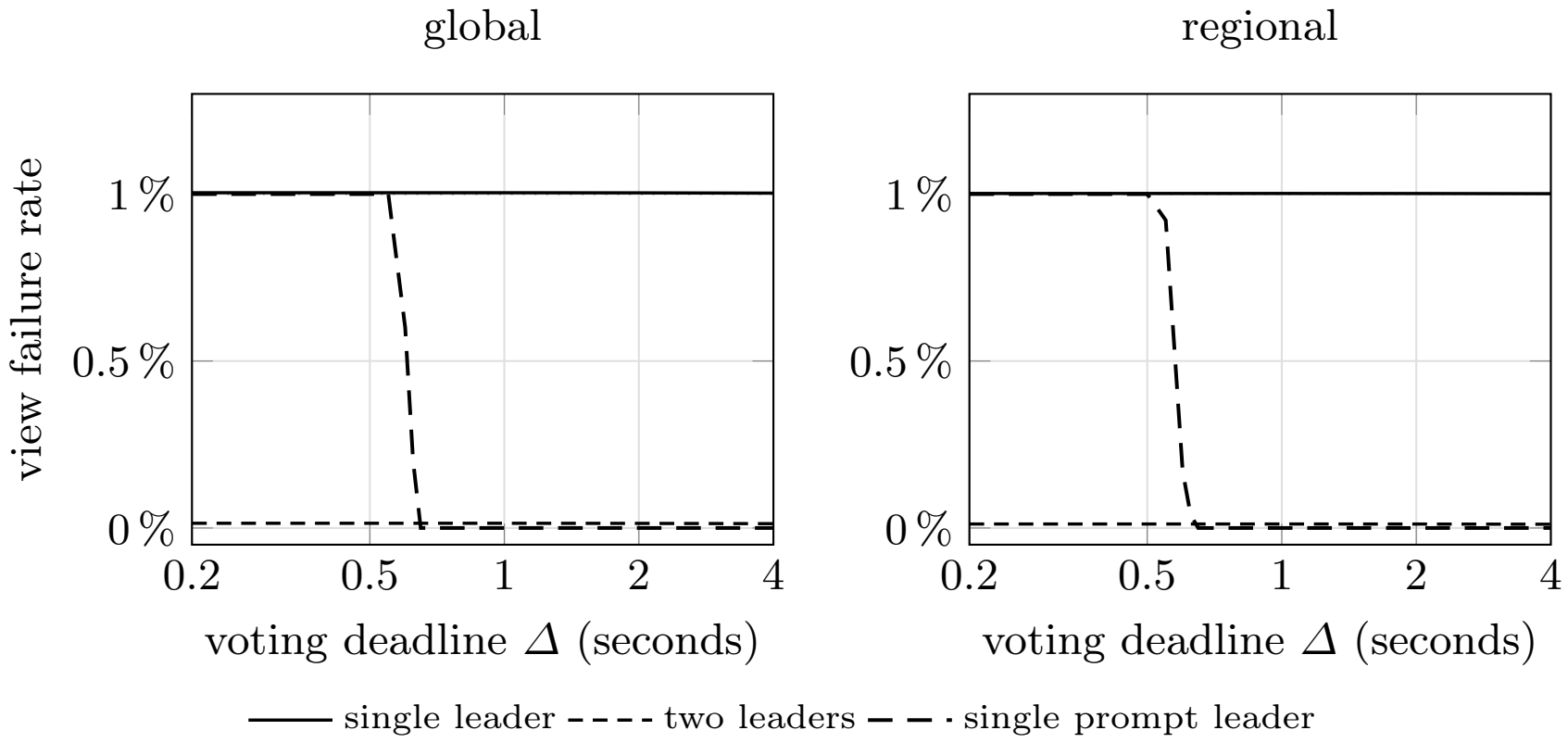

%% file: sections/related.tex
\section[Related Work]{Related Work}
\label{sec:related}

\paragraph{Consensus with multiple proposers.}
Many consensus protocols use multiple
proposers~\cite{bck24mysticeti,gnr25mcp,babel26cadence}.

Mir-BFT~\cite{sdpv21mir}
 distributes proposal work across parallel leaders.
MCP~\cite{gnr25mcp} combines proposals under one
leader for censorship resistance.
Ambulance~\cite{gmac25ambulance} uses a rigged race in which
the leader has a distinct advantage.
QuePaxa's asynchronous core~\cite{tbkk23quepaxa} uses private random proposal
priorities.
Its crash-fault model differs from our setting with
permanent adaptive
    Byzantine corruptions.
Gosig~\cite{li20gosig} also selects one proposal from a small random
leader set, using privately computed scores revealed with proposals.

\paragraph{A certified outcome for every proposer.}
Cadence~\cite{babel26cadence} implements a virtual leader
whose success
    in a slot depends on obtaining a certified outcome for every proposer.
When each proposer either completes its broadcast at slot
start or remains silent, an honest
    quorum can certify inclusion or exclusion and finalize in three message
    delays.
    A crash during the broadcast can instead divide the votes so that
    neither outcome has a quorum. One such proposer prevents the slot
    from completing through the fast path, even if every other
    proposer succeeds. \Cref{sec:cadence-crash} gives an execution with one
    mobile crash that forces this outcome after $\GST$. Cadence then starts
    fallback agreement for the same slot. In our good case, a completed
    broadcast
    from the highest-priority leader suffices for finalization.

\paragraph{Secret leader selection.}
Algorand~\cite{ghmvz17algorand} uses cryptographic sortition to let
participants privately learn their selection and reveal proof when
they send a message. Our leaders are public, but their
priority order remains unpredictable,
    including to Byzantine leaders, until honest beacon shares are
    released at the voting deadline.

\paragraph{Publicly ranked proposals.}
The multi-slot protocol built from Prefix
Consensus~\cite{xts26prefix} uses a public proposal
ranking.
An earlier proposer that equivocates or sends to only some
replicas can
    prevent later proposals from being committed in that slot. The protocol
identifies that proposer and gives it lower priority in
later
    slots.

\paragraph{Ordered leaders in DAG protocols.}
Mysticeti~\cite{bck24mysticeti} orders leader slots and
stops output at
    the first unresolved slot. Partial dissemination can leave a slot without
    the evidence required either to commit or to skip it, delaying later
    blocks even when their leaders have enough support. A later leader can
    resolve the earlier slot through the indirect decision rule.
Shoal++~\cite{arun25shoalplus} uses different rules: within
each DAG it
    evaluates one consensus instance at a time and chooses later leaders
    as that instance resolves. If a later leader causes the initial leader to
    be skipped, intermediate tentative leaders can also be skipped. Recent work
    on
Barnacle~\cite{danezis26barnacle} adapts the number of leader slots to
balance lower latency against delays caused
by slow leaders. These designs order multiple proposals, whereas our view forms one certificate for a
leader whose priority is revealed
    after the adversary has chosen which broadcasts to interrupt.

\paragraph{Leaderless and adaptive designs.}
ICC publicly ranks backup leaders~\cite{cdhp22icc} and
delays their proposals according to rank. VABA runs $n$
parallel protocol instances and randomly selects one after
    they have run~\cite{ams19vaba}. Antoniadis, Desjardins, Gramoli, Guerraoui,
and Zablotchi~\cite{adggz21leaderless} define a consensus protocol to be
\emph{leaderless} if every correct process decides in
every eventually synchronous execution allowing one process to be suspended per
round. In their message-passing model, a suspended process
    sends and receives nothing throughout a request--reply round of two message
    delays. Their BFT-Archipelago protocol tolerates one such suspension
    together
    with $f-1$ Byzantine processes, and terminates
    deterministically within $O(n)$ rounds in the worst
    case.

The classical mobile-fault lower bound rules out deterministic
consensus with one
    faulty sender per round, even under synchrony~\cite{sw89healer,as24mobileflp}.
    Gafni and Losa~\cite{gl24healer} study equivalent formulations.
    A crash can interrupt a broadcast after partial dissemination, which
    whole-round suspension excludes.

\paragraph{Incentives in consensus.}
Strategic block withholding in Bitcoin shows how
    participants can benefit from deviating from a consensus
    protocol~\cite{es14selfish,ssz16optimal,ckwn16instability}. In timing games,
    leaders can delay proposals to collect more valuable transactions, at the
    risk
    of missing the voting deadline. This tradeoff has been
    modeled for Ethereum~\cite{sst23time}. Other work studies how
    to
    encourage earlier proposals~\cite{abs25timing,ktbr26incentives}. Our
    evaluation
    finds fewer failed views with multiple leaders, even though leaders still benefit from delaying. Other work studies rational behavior in
distributed systems outside
    blockchains~\cite{%
  adgh06distributed,adh13leader,aacdmp05bar,wlacd11regret,hv16rational}.

%% file: sections/comparison-attacks.tex
\section[Mobile Crashes and an Interrupted Proposer]{Mobile Crashes and an
Interrupted Proposer}
\label{sec:comparison-attacks}

\paragraph{Classical lower bound.}
The classical lower bound states that any deterministic
consensus protocol
    with agreement and unanimity validity has a nonterminating execution under
    send omissions affecting one sender per step,
    even in synchrony~\cite{sw89healer}. A crash can interrupt a
    broadcast after messages have been sent to some recipients, and the adversary can later
    move to another replica. Mobile delays and crashes give related
    formulations and proofs~\cite{as24mobileflp,gl24healer}.

\paragraph{A mobile crash that delays Mysticeti.}
Mysticeti orders leader slots, each identifying a proposer
and a round.
    It directly skips a slot when $Q=2f+1$ blocks in the next
    round have no parent from that proposer. Direct commitment instead requires
$Q$ certificates for the proposal. Its output follows the
fixed slot
    order and stops at the first unresolved slot~\cite[Sections~II.C and
        III.B, Algorithms~1--3]{bck24mysticeti}.

Consider four replicas $A,B,C,D$, with $f=c=1$
and no Byzantine
    corruptions. In round $r$, let the proposer slots be ordered $D,A,B$,
with $D$ the primary proposer. Replica $A$ sends its proposal only to $B$
before crashing.
    Keep $A$ crashed until the other replicas have constructed their blocks
    for rounds $r+1$ and $r+2$. Arrange delivery so that $B$
    knows $A$'s
    proposal when it creates its round-$r+1$ block, whereas $C$
    and $D$ create
    their round-$r+1$ blocks before learning that proposal. Deliver $B$'s
proposal to all three active replicas before their voting
blocks are
    created. The resulting references are
\[
    \begin{array}{c@{\quad}|@{\quad}c@{\quad}c}
        \text{round-$(r+1)$ block} & \text{supports }A & \text{supports }B \\
        \hline
        B_{r+1}                    & \text{yes}        & \text{yes}        \\
        C_{r+1}                    & \text{no}         & \text{yes}        \\
        D_{r+1}                    & \text{no}         & \text{yes}.
    \end{array}
\]
These blocks are valid: each can reference the three
round-$r$ blocks of
    $B,C,D$, and $B_{r+1}$ also references $A$'s block. Deliver
    the three voting blocks and their dependencies, and let each active
    replica's round-$r+2$ block reference all three. Every such block certifies
$B$, while none certifies $A$. Thus $B$
has a direct-commit certificate,
    but $A$ has only one supporting voting block and two that do not support it.
    Neither count reaches three, so $A$ is neither directly committed nor
    directly skipped.

Mysticeti waits for the primary proposer, here $D$. Let $B,C,D$ enter the proposal
    round at time $0$ after $\GST$, and let $A$ enter at
    $\Delta/2$. Deliver $A$'s proposal to
$B$ at $3\Delta/4$.
    Replicas $C,D$ create their voting blocks at $\Delta/4$, after receiving the proposals of $B,C,D$.
    Replica $B$ receives $C$'s proposal at $\Delta/4$ and $D$'s
    at $\Delta$, so it votes after receiving $A$'s proposal.
    Delay $C,D$'s voting blocks to $B$ until $5\Delta/4$ so they do not
    reveal $D$'s proposal earlier.
     Later delivery provides the dependencies needed to
    validate the voting blocks.

Mysticeti's indirect rule considers only slots in rounds
strictly greater
    than the unresolved slot's decision round, which is $r+2$ for its default
    three-round configuration. Hence a later slot in round $r$ cannot resolve
    it.
    Algorithm~3 stops ordered output at $A$, even though $B$
    is already
    marked for commitment. A resolving proposal must come from round $r+3$
    or later.

%% file: sections/cadence-attack.tex
\subsection[A Mobile Crash That Forces Cadence to Use Fallback]{A Mobile
Crash That Forces Cadence to Use Fallback}
\label{sec:cadence-crash}

Cadence requires a quorum of matching
inclusion or exclusion votes for
    each proposer before its fast path can commit the
    slot~\cite[Section~4.2 and Algorithms~2--4]{babel26cadence}.
    The execution below uses one immediate crash, a budget refund delay $\tau=\Delta$,
    and no Byzantine corruption.

Consider four replicas $P,A,B,C$, with
$f=c=1$, and two honest proposers
    $P,A$. The slot starts at time $0$ after $\GST$, and its
    proposal deadline
    is $\Delta$. Fix $0<\varepsilon<\Delta/4$.
\begin{enumerate}
    \item $P$ sends its assigned chunk to itself and the
    chunk assigned to
              $A$, then crashes before sending to $B,C$.
    \item Heal $P$ at time
    $\varepsilon$. It resumes sending to $B,C$.
              Deliver these messages at $\Delta+\varepsilon$, a delay of
              exactly
          $\Delta$. Deliver $A$'s chunk from
          $P$ at $\Delta-\varepsilon$, and the other proposer's chunks
              to their recipients, by $\Delta-\varepsilon$, with
              $P$'s
              chunk from $A$ arriving at that time.
    \item By the deadline, $P,A$ vote to include $P$'s
    proposal, while
              $B,C$ vote to exclude it. Any chunks forwarded after reception at
              $\Delta-\varepsilon$ can arrive after the deadline while
              respecting
              the network delay bound. This also covers early voting.
              All four replicas vote to include $A$'s proposal.
    \item Neither set of two votes for $P$
    reaches the quorum of three.
          Chunks received later do not change the signed
          votes. Thus, no certified outcome
              for $P$ can be formed by this protocol, and
              the slot requires fallback.
\end{enumerate}

Cadence's fallback starts a separate multivalued Byzantine
    agreement protocol for the same slot~\cite[Section~4.3]{babel26cadence}.